\documentclass{llncs}
\usepackage[german,english]{babel}
\usepackage{times,theorem}
\usepackage{latexsym,color,graphics}
\usepackage{diagrams1}
\diagramstyle[tight,centredisplay%
,dpi=600
,UglyObsolete
,heads=LaTeX%
]

\usepackage{epsfig}
\usepackage{amssymb}
\usepackage{amsmath}
\usepackage{amsfonts}
\usepackage{xspace}
\usepackage{graphicx}
\begingroup
\catcode`\~=11
\gdef\urltilde{\lower 0.6ex\hbox{~}}
\endgroup

\newcommand{\A}{\mathcal{A}} \newcommand{\B}{\mathcal{B}}
 \newcommand{\D}{\mathcal{D}}
\newcommand{\E}{\mathcal{E}} 
 
\newcommand{\I}{\mathcal{I}} 
 \renewcommand{\L}{\mathcal{L}}
\newcommand{\M}{\mathcal{M}} \newcommand{\N}{\mathcal{N}}
 \renewcommand{\P}{\mathcal{P}}
 \newcommand{\R}{\mathcal{R}}
\renewcommand{\S}{\mathcal{S}} \newcommand{\T}{\mathcal{T}}
 \newcommand{\V}{\mathcal{V}}
\newcommand{\W}{\mathcal{W}}

\title{Intensional FOL over Belnap's Billatice for Strong-AI Robotics}
\author{Zoran Majki\'c}
\authorrunning{Zoran Majki\'c}
\institute{ISRST, Tallahassee, FL, USA\\
\email{majk.1234@yahoo.com}}

\newtheorem{theo}{Theorem}

\newtheorem{coro}{Corollary}

\begin{document}
\maketitle              

\begin{abstract}
 AGI (Strong AI) aims to create intelligent robots that are quasi indistinguishable from the human mind. Like a child, the AGI robot would have to learn through input and experiences, constantly progressing and advancing its abilities over time. The AGI robot would require an intelligence more close to human's intelligence: it would have a self-aware consciousness that has the ability to solve problems, learn, and plan.\\
Based on this approach  an Intensional many-sorted First-order Logic (IFOL), as an extension of a standard FOL with Tarskian's semantics, is proposed in order to avoid the problems of standard 2-valued FOL with paradoxes (inconsistent formulae) and a necessity for robots to work with incomplete (unknown) knowledge as well. This is a more sophisticated version of IFOL with the same syntax  but different semantics, able to deal with truth-ordering and knowledge-ordering as well, based on the well known Belnap's billatice with four truth-values that extend the set of classical two truth-values.
\end{abstract}
Keywords: AGI, Belnap's bilattice, First-order Logic, Robotics

\section{Introduction}
\begin{figure}
$\vspace*{-9mm}$
\centering{
 \includegraphics[scale= 1.11]{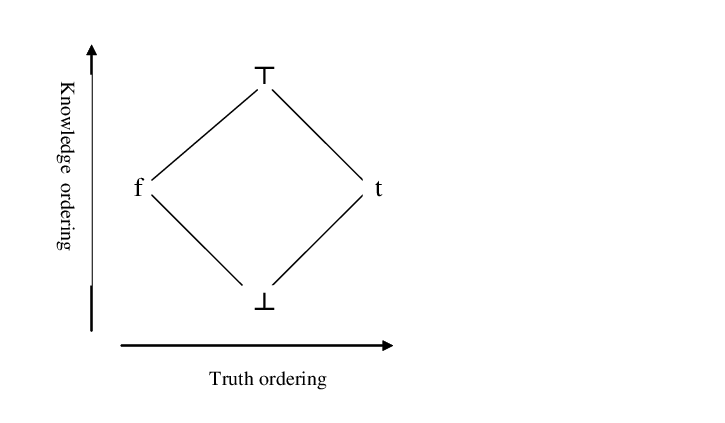}
  \caption{Belnap's bilattice}
  \label{fig:Operations}
 }
  $\vspace*{-11mm}$
 \end{figure}

Many-valued logic was conceived as a logic for uncertain, incomplete
and possibly inconsistent information which is very close to the
statements containing the words "necessary" and "possible", that is,
to the statements that make an assertion about the \emph{mode of
truth} of some other statement. \emph{Algebraic} semantics interprets modal connectives as operators, while \emph{Relational} semantics uses
relational structures, often called Kripke models, whose elements
are thought of variously as being possible worlds; for example,
moments of time, belief situations, states of a computer, etc.. The
two approaches are closely related: the subsets of relational
structures form an algebra with modal operators, while conversely
any modal algebra can be embedded into an algebra of subsets of a
relational structure via extensions of Stone's
representation theory.\\
 In ~\cite{Beln77} Belnap introduced the 4-valued
bilattice, $X= \B_4 = \{t,f,\bot, \top\}$  in Fig.\ref{fig:Operations}, with four truth-values in $X$ where $t $ is \emph{true}, $f $ is \emph{false}, $\top$ is inconsistent (both true and false) or \emph{possible }, and $\bot$ is \emph{unknown} truth-value. In what follows  we denote by $x \bowtie y$ two unrelated elements in $X$ (so that not $(x\leq y$ or $y\leq x)$). So, Belnap's bilattice is composed by the truth lattice $(X,\leq)$ and knowledge lattice $(X,\leq_k)$,
with two natural orders:
\emph{truth} order, $\leq$, and \emph{knowledge} order, $\leq_k$,
such that $f \leq \top \leq t$, $~f \leq \bot \leq t$, $\bot
\bowtie_t \top$ and $\bot \leq_k f \leq_k \top$, $~\bot \leq_k t
\leq_k \top$, $f \bowtie_k t$. That is, bottom element for
$\leq$ ordering is $f$, and for $\leq_k$ ordering is $\bot$,
 and top element for $\leq$ ordering is $t$, and for $\leq_k$ ordering is $\top$. Meet and join operators under $\leq$ are denoted $\wedge$ and
$\vee$; they are natural generalizations of the usual conjunction
and disjunction notions. Meet and join under $\leq_k$ are denoted
$\otimes$ and $\oplus$, such that hold: $~f \otimes t = \bot$, $f
\oplus t =\top$, $\top\wedge \bot = f$ and $\top \vee \bot = t$.

The truth lattice will be used for the many-valued many-sorted intensional FOL (denoted by $IFOL_B$) while the knowledge lattice can by used for estimation of the rate of growing the robots knowledge in time, as result of robot's experience and deductions. In order to simplify presentation we will use the same symbols of logic connectives of $IFOL_B$ as that of Belnap's bilattice for conjunction, disjunction, implication and negation.
So, the conjunction and disjunction of $IFOL_B$ are consequently the meet and join operators under $\leq$, denoted $\wedge$ and
$\vee$.  For the implication $\Rightarrow$ of  $IFOL_B$ we can use the \emph{relative pseudocomplements}, defined for the truth values by by $x \Rightarrow y = \bigvee\{z~|~z \wedge x \leq y\}$, while for the negation, denoted $\neg$, the \emph{paraconsistent} negation(reverses the $\leq$ ordering, while preserving the $\leq_k$ ordering)\footnote{Differently from it, the negation used as pseudocomplement, $\neg_t x = x \Rightarrow f$, reverses both orderings.}: switching $f$ and $t$, leaving $\bot$ and $\top$.
\\ The many-valued equivalence operator is defined for any two $a,b \in X$ by:
\begin{equation} \label{eq:dueM2}
a \Leftrightarrow b =  \left\{
    \begin{array}{ll}
      t, & \hbox{if $~a = b$}\\
      f, & \hbox{otherwise}
    \end{array}
  \right.
\end{equation}
 In what follows we will consider only conservative homomorphic
many-valued extensions of the classic 2-valued logic with logical
connectives $\neg, \wedge, \vee,  \Rightarrow,\Leftrightarrow$
(negation, conjunction, disjunction, implication and equivalence
respectively) and of the following algebra of truth values in the lattice $(X,\leq)$:
\begin{definition} \label{def:mv-algebra}
 For a given many-valued predicate logic,  by $\A_{B} = (X, \leq, \neg, \wedge, \vee,  \Rightarrow,\Leftrightarrow)$ we denote the algebra of the Belnap's truth values in the complete lattice $(X, \leq)$.
\end{definition}
The fact that classic 2-valued algebra is the subalgebra of this
many-valued algebra $X$, means that $X$ is a conservative
extension of the classic logic operators.\\
\textbf{Remark}: It has been demonstrated by Lemma 1 in  \cite{Majk07MV} that the unary operator of negation $\neg$ in the Belnap's algebra $\A_B$ is the selfadjoint modal operator w.r.t the knowledge-ordering $\leq_k$, and antitonic auto-homomorphism of the truth-ordering lattice, $\neg:(X,\leq,\wedge,\vee) \rightarrow (X,\leq,\wedge,\vee)$, which represents the De Morgan laws.
 So, the algebra of truth values $\A_B$ is a \emph{modal} Heyting algebra (that is, a standard Heyting algebra extended by unary modal operator).\\
$\square$\\
 For the existential and universal quantifiers of predicate logic $IFOL_B$ with the set of variables $\V$ and domain $\D$ (that used of intensional FOL \cite{Majk22}) the semantics will be defined in what follows.

 An assignment $g:\V \rightarrow \D$ for variables in $\V$  in  $IFOL_B$ is
applied only to free variables in terms and formulae.
 Such an assignment $g \in \D^{\V}$ can be recursively uniquely extended into the assignment $g^*:\T \rightarrow \D$, where $\T$ denotes the set of all terms, by:
\begin{enumerate}
  \item $g^*(t_i) = g(x) \in \D$ if the term $t_i$ is a variable $x \in \V$.
  \item If the term $t_i$ is a constant $c \in F$ then $g^*(t_i)  \in \D$ us its tarskian interpretation.
  \item If a term $t_i$ is $f_i^k(t_1,...,t_k)$, where $f_i^k \in F$ is a
k-ary functional symbol and $t_1,...,t_k$ are terms, then
$g^*(f_i^k(t_1,...,t_k))$ is the value $u\in \D$ of this functions for the tuple of values in  $(g^*(t_1),...,g^*(t_k))$ or,
equivalently, in the graph-interpretation of the function,
  $(g^*(t_1),...,g^*(t_k),u) \in \D^{k+1}$.
\end{enumerate}
In what follows we will use the graph-interpretation for functions
 in FOL like its interpretation in intensional logics. We denote by $\L$ the set of all formulae $\phi$ of the logic $IFOL_B$, and denote by $~t_i/g~$ (or $\phi/g$) the ground term (or formula) without free variables, obtained by assignment $g$ from a term $t_i$ (or a formula $\phi$), and by  $\phi[x/t_i]$ the formula
obtained by  uniformly replacing $x$ by a term $t_i$ in open formula $\phi(x)$.\\
A \emph{sentence} is a (closed) formula having no free variables.
%

 We denote by $\textbf{2} = \{f,t\}\subseteq X $ the
classical 2-valued logic lattice, which is the sublattice of the Belnap's truth-ordering lattice $(X, \leq)$ composed only by false and true truth-values $\{f,t\}$. It is easy to verify that Definition \ref{def:HerbrandValuat} is valid also for classical FOL and its 2-valued logic lattice $\textbf{2}$. However, the Belnap's bilattice has also the knowledge ordering which is very useful for estimation of the robot's knowledge, because for a robot, as default any ground atom has the unknown truth-value $\bot$.

 In what follows any open-sentence, a formula $\phi(\textbf{x})$ with non empty tuple of free variables $\textbf{x} =(x_1,...,x_m)$, will be called a m-ary   \emph{virtual predicate}\footnote{In intensional FOL we use virtual pradicates so that their extension for a given interpretation generates a single m-ary relation as in the case of standard m-ary predicates used in FOL for atomic formulae with $m\geq 1$ free variables. Each virtual predicate generates a m-ary \emph{intensional concept} in domain $\D$ for a given intensional interpretation as standard m-ary predicates. Most simple syntactically virtual predicates are obtained from atoms, obtained from k-ary predicate letters $p^k_i \in P$, with non empty set of variables and non empty set of constants for predicate arguments as, for example $p^6_i(t_1,t_2,x_1,t_3,x_2,t_4)$ with ground terms (or constants), $t_m$ for $1\leq m\leq 4$, generates a virtual predicate $\phi_i(x_1,x_2)$. Virtual predicate of an atom with all its arguments represented by variables is just equal to such atom.}, denoted also by $\phi(x_1,...,x_m)$. This definition contains the precise method of establishing the \emph{ordering} of variables in this tuple:
\begin{definition} \label{def:virt-predicate} \textsc{Virtual predicates:}
\emph{Virtual predicate} obtained from an open formula $\phi \in \L$
is denoted by $\phi(x_1,...,x_m)$ where $(x_1,...,x_m)$ is a particular fixed sequence of the set of all free variables in $\phi$. This definition contains the precise method of establishing the \emph{ordering} of variables in this tuple:
such an method that will be adopted here is the ordering of appearance, from left to right, of free variables in $\phi$. This method of composing the tuple of free variables is  unique and canonical way of definition of the virtual predicate from a given open formula.
\end{definition}
Virtual predicates are used to build the \emph{semantic logic structures} of logic-semantics level of any given natural language. However, with virtual predicates we need to replace the general FOL quantifier on variables $(\exists x)$ by specific existential quantifiers $\exists_i$ of the intensional logic $IFOL_B$, where $i\geq 1$ is the position of variable $x$ inside a virtual predicate. For example, the intensional FOL formula $(\exists x_k) \phi(x_i,x_j,x_k,x_l,x_m)$ will be mapped into intensional concept $\exists_3 \phi(\textbf{x})$ where $\textbf{x}$ is the list(tuple) of variables $(x_i,x_j,x_k,x_l,x_m)$.
In the same way we introduce the set of universal quantifiers $\forall_i$.  Notice that in many-valued version of intensional FOL the composed operation $\neg \exists_i \neg$ \emph{is different} from $\forall_i$. Moreover with virtual predicates we need to replace standard binary logic connectives $\wedge, \vee, \Rightarrow$ with $\wedge_S, \vee_S, \Rightarrow_S$ where $S$ is a set of pair of indices of variables in virtual predicate. So we obtain a non standard FOL syntax algebra with such expressions that use these new connectives.

The significant aspect of an expression's meaning is its \emph{extension}.
We can stipulate that the extension of a sentence is its
truth-value, and that the extension of a singular term is its
referent. The extension of other expressions can be seen as
associated entities that contribute to the truth-value of a sentence
in a manner broadly analogous to the way in which the referent of a
singular term contributes to the truth-value of a sentence.

The first conception of \emph{intensional entities} (or concepts) is built
into the \emph{possible-worlds} treatment of Properties, Relations
and Propositions (PRP)s. This conception is commonly attributed to
Leibniz, and underlies Alonzo Church's alternative formulation of
Frege's theory of senses ("A formulation of the Logic of Sense and
Denotation" in Henle, Kallen, and Langer, 3-24, and "Outline of a
Revised Formulation of the Logic of Sense and Denotation" in two
parts, Nous,VII (1973), 24-33, and VIII,(1974),135-156). This
conception of PRPs is ideally suited for treating the
\emph{modalities} (necessity, possibility, etc..) and to Montague's
definition of intension of a given virtual predicate
$\phi(x_1,...,x_k)$, as a mapping from possible worlds into
extensions of this virtual predicate. Among the possible worlds we
distinguish the \emph{actual} possible world\footnote{For example, if we
consider a set of predicates, of a given Database,
and their extensions in different time-instances, then the actual possible world is identified by the current instance of the time.}.

The second conception of intensional entities is to be found in
Russell's doctrine of logical atomism. In this doctrine it is
required that all complete definitions of intensional entities be
finite as well as unique and non-circular: it offers an
\emph{algebraic} way for definition of complex intensional entities
from simple (atomic) entities (i.e., algebra of concepts),
conception also evident in Leibniz's remarks. In a predicate logics,
predicates and open-sentences (with free variables) expresses
classes (properties and relations), and sentences express
propositions. Note that classes (intensional entities) are
\emph{reified}, i.e., they belong to the same domain as individual
objects (particulars). This endows the intensional logics with a
great deal of uniformity, making it possible to manipulate classes
and individual objects in the same language.

After publication of my book \cite{Majk22} in 2022 for such IFOL (Intensional FOL), recently I applied the standard 2-valued version of it \cite{Majk23r,Majk24a,Majk24ar} to the four-level cognitive structure of AGI robots here adopted for Belnap's 4-valued bilattice that requires the different structure of the conceptual PRP structure levels.

Consequently, for this particular application of intensional FOL to the strong-AI robots, for the extensional representation of the concepts in this new logic $IFOL_B$, differently from the general many-valued intensional first-order logics, provided in  Section 5.2 in \cite{Majk22}, the rule of \emph{false statements} (non represented explicitly per default in concepts extensions) will be replaced \emph{by unknown} statements. So, here in next section we need to redefine these general many-valued definitions, used for the algebras of concepts and extensions, by new definitions for this new logic $IFOL_B$ based on Belnap's bilattice in Fig.\ref{fig:Operations}.
\section{Belnap's 4-valued Many-sorted Intensional First-order Logic Syntax with Abstraction Operator\label{sec:MVFOL}}
The basic issue of the work in this section is a definition of such a new logic $IFOL_B$ based on Belnap's bilattice, by generalization of the standard  FOL,
able to distinguish the intensional and extensional aspects of the
semantics and to deal with incomplete and contradictory information, able as the human minds to deal with Lear paradoxes as well. The distinction between intensions and extensions is important, considering that extensions can be
notoriously difficult to handle in an efficient manner.\\
 However, differently from 2-valued logic where extension of a given predicate is composed by only \emph{true} atoms, here for MV-logic $IFOL_B$ the ground atoms of extension of this predicate can have three truth-values (\emph{except the unknown value} $\bot$ for unknown facts which are not explicitly represented). In this way, the complete knowledge of AGI robots would correspond to the extensions of the only known concepts and hence it would be very easy to compare the knowledge of different robots developed by their individual experiences supported by their deductive inference capabilities.

 The syntax of the many-valued intensional first-order logic extends
 the syntax of the standard two-valued First-order Logic (FOL) with identity, enriched by abstraction operator $\lessdot \_~\gtrdot$ provided in Section 1.3.1 in \cite{Majk22} and more precisely and recently in many-sorted version as is $IFOL_B$ by Definition 4 in  \cite{Majk24ar}, and hence
 with extended set of terms obtained by the abstraction of logic \index{Intensional Many-valued FOL} formulae. However, in our case we have no Tarskian interpretations of this logic by a many-valued (based on Belnap's 4-valued bilattice) interpretations, and hence here we provide a shortened version in next definition.

  We recall that we denote by $~t/g~$ (or $\phi/g$) the ground term (or
formula) without free variables, obtained by assignment $g$ from a
term $t$ (or a formula $\phi$), and by  $\phi[x/t]$ the formula \index{intensional abstraction} obtained by  uniformly replacing $x$ by a term $t$ in $\phi$.
\begin{definition} \label{def:abstrConv} \textsc{Intensional abstraction convention}:

 From the fact that we can use any permutation of the variables in a given virtual predicate,  we introduce the convention that
 \begin{equation}\label{eq:abstrctConv}
 \lessdot \phi(\textbf{x})\gtrdot_{\alpha}^{\beta}~~ is~a~ term~ obtained~ from~ virtual ~ predicate ~~\phi(\textbf{x})
 \end{equation}
 if $\alpha$ is \textsl{not empty}   such that  $\alpha\bigcup\beta$ is  the set of all variables in the list (tuple of variables)  $\textbf{x} = (x_1,...,x_n)$ of the virtual predicate (an open logic formula) $\phi$,  and $\alpha\bigcap\beta = \emptyset$, so that $|\alpha|+|\beta| = |\textbf{x}| = n$.
 Only the variables in $\beta$ (which are the only free variables of this term), can be quantified. If $\beta$ is empty then $\lessdot \phi(\textbf{x})\gtrdot_{\alpha}$ is a \emph{ground term}. If $\phi$ is a sentence and hence both $\alpha$ and $\beta$ are empty, we write simply $\lessdot \phi \gtrdot$ for this ground term.

The abstracted terms $\lessdot \phi \gtrdot_\alpha^\beta$ can be used as terms in any predicate $p_j^k\in P$ for reification purpose, for example for an atom $p^3_j(\lessdot\phi \gtrdot_\alpha^\beta,y,z)$ with free variables $y$, $z$ and that in $\beta$.
Let $p^k_j(t_1,...,t_k)$ be an atom  with at least one of abstract term $t_i = \lessdot \phi_i(\textbf{x}_i)\gtrdot_{\alpha_i}^{\beta_i}$ with $\beta_i$ non empty and let $\beta$ denotes the union of all $\beta_i$ of the abstracted terms in this atom. We can consider this atom  as a virtual predicate $\phi(\textbf{x})$ with ordered tuple of free variables $\textbf{x}$, and  we denote by $\textbf{y}$ its ordered subtuple  without variables in $\beta$ with $1\leq n= |\textbf{y}| $.  Then we have that for each assignment $g \in \D^\beta$, $p^k_j(t_1/g,...,t_k/g)$ is a standard atom (all abstracted terms $t_i/g = g^*(t_i)$ by using (\ref{eq:assAbTerm}) are transformed to values in $\D$.
\end{definition}

  In what follows, this logic $IFOL_B$  will be shortly denoted by $\L_{in}$ and the set of all its formulae by $\L$.
 \begin{definition} \label{def:syntax} \textsc{Syntax of many-valued intensional FOL $\L_{in}$}:\\
 We define the syntax of 4-valued many-sorted first-order logic $\L_{in}$,  with complete Belnap's lattice of truth-values $(X,\leq)$, by:\\
 - Variables $x,y,z,..$ in $\V$;\\
 - Language constants $c,d,...$ are considered as nullary functional letters (all non nullary functional letters are represented as the predicate
  letters for graphs of these functions);  \\
  - Predicate letters in $P$, denoted by $p_1^{k_1},p_2^{k_2},...$
   with a given arity $k_i\geq 1$,    $i = 1,2,..$  Nullary predicate letters are considered as propositional letters, as for example the built-in predicate letters $\{p_a~|~ a \in X\}$;\\
  - The 4-valued logic connectives:
   1. unary connectives are the negation $\neg$, and existential an universal quantifiers $\exists_i, \forall_i$.\\
   2. binary connectives in $\circledcirc \in \{\wedge,\vee,\Rightarrow, \Leftrightarrow \}$ are   respectively logic conjunction, disjunction,  implication and  equivalence.
   From the fact that we are using the virtual predicates that represent any formula composed by these connectives as a specific virtual predicate with free variables obtained as the union of the free variables of all its standard predicates in $P$ that compose this formula, for the syntax algebra $\mathcal{A}\mathfrak{B}_{FOL}$ (in next Corollary \ref{coro:intensemantMV} ) of $\L_{in}$ we need to use for each binary logic operator in $\circledcirc$ (different from $\Leftrightarrow$) an algebraic operator $\circledcirc_S$ where the pairs of indexes in the set $S$ indicate the equal free variables used in two subformulae $\psi_1$ and $\psi_2$ that compose a resulting virtual predicate $\phi(x)$ equal to formula $\psi_1\circledcirc_S \psi_2$ as it will be explained\footnote{
  For example, the FOL formula
$\psi_1(x_i,x_j,x_k,x_l,x_m) \wedge \psi_2 (x_l,y_i,x_j,y_j)$ will be
replaced by a resulting \emph{virtual predicate} $\phi(x_i,x_j,x_k,x_l,x_m,y_i,y_j)$ defined by algebraic expression $\psi_1(x_i,x_j,x_k,x_l,x_m)
\wedge_S \psi_2~ (x_l,y_i,x_j,y_j)$, with $S = \{(4,1),(2,3)\}$, and then traduced by
the algebraic expression $~R_1 \bowtie_{S}R_2$  (see point 2.5 of Definition \ref{def:m-extensions}) where $R_1 \in
\P(\D^5), R_2\in \P(\D^4)$ are the extensions for a given many-valued
interpretation $v^*$ of the virtual predicate $\psi_1, \psi_2$
relatively. In this example the resulting relation
will have the following ordering of attributes of obtained virtual predicate $\phi$:
$(x_i,x_j,x_k,x_l,x_m,y_i,y_j)$.\\
In the case when $S$ is empty (i.e. its cardinality $|S| =0$) then the resulting relation is the Cartesian product of $R_1$ and $R_2$.}
   in what follows.\\
  - Abstraction operator $\lessdot \_~\gtrdot$, and punctuation
 symbols (comma, parenthesis).\\
 With the following simultaneous inductive definition of \emph{terms} and \emph{formulae}:
 \begin{enumerate}
   \item All variables and constants (0-ary functional letters in P) are terms.
   We denote by $p_a$ a logic constant (built-in 0-ary predicate symbol) for each truth value $a \in X$, so that a set of constants is a not empty set.
   \item If $~t_1,...,t_k$ are terms, then $p_i^k(t_1,...,t_k)$ is a formula
 ($p_i^k \in P$ is a k-ary predicate letter).
   \item  In what follows any open-formula $\phi(\textbf{x})$ with non empty
tuple of free variables $\textbf{x} = (x_1,...,x_m)$, will be called a m-ary
  \emph{virtual predicate}, denoted also by
$\phi(x_1,...,x_m)$ and provided in Definition  \ref{def:virt-predicate}.\\
   If $\phi$ and $\psi$ are formulae, then   $\neg\phi$, $(\forall_k x) \phi$,  $(\exists_k x) \phi$ and, for each connective $\odot$, if binary $\phi \odot \psi$ or if unary $\odot\phi$, are the formulae. In a formula $(\exists_k )\phi(\textbf{x})$ (or $(\forall_k) \phi(\textbf{x})$), the virtual predicate $\phi(\textbf{x})$ is called "action field" for the quantifier $(\exists_k)$ (or $(\forall_k)$) of the k-th free variable in the tuple $\textbf{x}$. A variable $y$ in a formula $\phi$ is called bounded variable iff it is the variable quantified by $(\exists_k)$ (or $(\forall_k)$). A variable $x$ is
free in $\psi(\textbf{x})$ if it is not bounded. A \emph{sentence} is a closed-formula having no free variables.
\item If $\phi(\textbf{x})$   is a formula and $\alpha \subseteq \overline{\textbf{x}}$ is a possibly empty subset of  \emph{hidden} (compressed) variables, then $\lessdot \phi(\textbf{x}) \gtrdot_{\alpha}^{\beta}$ is an abstracted term, as provided in Definition \ref{def:abstrConv},  where $\beta$ is  remained subset of free visible variables in  $\phi$.  So, the subtuples of hidden and visible variables (preserving the ordering of the tuple $\textbf{x}$ are $\pi_{-\beta}\textbf{x}$ and $\pi_{-\alpha}\textbf{x}$, respectively).  If $\alpha$ or $\beta$ is empty sequence, than it can be
 omitted (for example, if $\phi$ is closed formula, then this term
 is denoted by $\lessdot \phi \gtrdot$).
  \end{enumerate}
An occurrence of a variable $x_i$ in  a term $\lessdot \phi(\textbf{x})
\gtrdot_{\alpha}^{\beta}$ is \emph{bound}  if $x_i \in \alpha$, free
if $x_i \in \beta$, so that the variables in $\alpha$ are not subjects of assignment $g\in \D^\V$ and can not be quantified by existential and universal FOL quantifiers.\\
 In particular we introduce the following distinguished predicates:
\begin{description}
 \item[a] The binary predicate letter $p_1^2 \in P$ is singled out as a distinguished
logical predicate and formulae of the form $p_1^2(t_1,t_2)$ are to
be rewritten in the form $t_1 = t_2$. It is a 2-valued built-in
predicate for the identity.
  \item[b] The binary predicate letter $p_2^2 \in P$ is singled out as a
distinguished logical predicate and formulae of the form
$p_2^2(t_1,t_2)$ are to be rewritten in the form $t_1 =_{in} t_2$.
It is a 2-valued built-in
predicate for the weak-intensional-equivalence.
  \item[c] The unary predicate letter $p_1^1 \in P$ is singled out as
distinguished many-valued truth-predicate for sentences so that the sort of its argument is that of abstracted terms. Ground atoms of the form $p_1^1(t_1)$, where the term $t_1$ is abstracted term obtained by reification of a sentence $\phi(\textbf{x})/g$,  are to be rewritten in the form $T(t_1)$ with constraint that the logic truth-value of $T(\lessdot\phi(\textbf{x})/g)\gtrdot)$ is imposed to be equal to the truth-value of the sentence $\phi(\textbf{x})/g$.
\end{description}
\end{definition}
Notice that "built-in" is used for predicate letters that have fixed
invariant interpretation, and consequently fixed invariant
extension: each possible interpretation of $\L_{in}$ has to satisfy
this constraint for built-in predicates, so that different many-valued
interpretations can be used only for the remaining set of predicate
letters in $P$.
\begin{definition} \label{def:HerbrandValuat}
 The Herbrand base of a  logic $IFOL_B$ is defined by

 $~~H = \{p_i^k(t_1,..,t_k)~|~p_i^k \in P$ and $t_1,...,t_k$ are ground terms $\}$.
 \\
 Herbrand interpretations  are the mappings $v:H \rightarrow X$,which must satisfy the constraints for the  built-in ground atoms (any ground atom of a built-in predicates must have the same truth-value for every Herbrand interpretation) and for built-in propositional symbols $p_a$, for $a \in X$, $v(p_a) =a$.
\end{definition}
 Note that  if we desire to control if, for a given interpretation $v$,  $a\in X$ is the truth-value of a sentence $\phi$, than it is enough to control if formula $p_a\Leftrightarrow \phi/g$ is true. Consequently, for any sentence $\phi(\textbf{x})/g$ the 2-valued logic truth-scheme formulae for any truth-value $a \in X$
\begin{equation}\label{eq:truthScheme}
p_a\Leftrightarrow \phi(\textbf{x})/g
\end{equation}
has distinguished importance for reduction of many-valued sentences to 2-valued formulae that can be only true or false.

We are able to define the many-valued algebraic semantics of $\L_{in}$,
based on the standard extension of Herbrand interpretations to all
sentences $\L_0$ of $\L_{in}$.
\begin{definition}  \label{def:MV-algebra} \textsc{Belnap's 4-valued Semantics of $\L_{in}$}:\\
The algebraic semantics of the 4-valued many-sorted intensional first-order logic $\L_{in}$ given in Definition \ref{def:syntax}
can be obtained by the unique extension of a given 4-valued Herbrand interpretation $v:H\rightarrow X$ (from Definition \ref{def:HerbrandValuat}), into the valuation  $v^*:\L_0 \rightarrow X$, where $\L_0 \subset \L$ is the strict subset of all sentences
(formulae without free variables), inductively as follows: \\for any formula $\phi, \psi \in \L$ and a given assignment $g:\V \rightarrow \D$,  we have that (here $\bigwedge$ and $\bigvee$ are the meet and joint operators of the lattice $X$, respectively)
\begin{enumerate}
  \item $~~v^*(\neg \phi/g) = ~\neg v^*(\phi/g)$, and for built-in propositions $v^*(p_a) = v(p_a)$ for $a\in X$,
  \item $~~v^*(\phi/g \odot \psi/g) =  v^*(\phi/g) \odot
v^*(\psi/g)$, \\for each logic connective $\odot \in \{\wedge, \vee,
\Rightarrow, \Leftrightarrow\}$, i.e., $\A_{B}$-operators in Definition \ref{def:mv-algebra} over Belnap's truth-lattice $X$,
  \item $~~v^*(((\exists_i) \phi)/g) =  v^*(\phi/g)$ if i-th variable $x$ is not a free  in $\phi$;\\
 $ = \bigvee \{v^*(\phi/g_1)~|~g_1 \in \D^{V}$ such that for all $y \in \V \backslash\{x\}, g_1(y) = g(y)\}$ otherwise,
  \item $~~v^*(((\forall_i) \phi)/g) =  v^*(\phi/g)$ if i-th variable $x$ is not a free  in $\phi$;\\
 $ = \bigwedge \{v^*(\phi/g_1)~|~g_1 \in \D^{V}$ such that for all $y \in \V \backslash\{x\}, g_1(y) =  g(y)\}$ otherwise.
\end{enumerate}
The distinguished unary predicate $T$  in point $\textbf{c}$ of the syntax definition above is the many-valued version of the truth-predicate in the standard 2-valued logic where a formula $T(\lessdot\phi/g(\textbf{x})/g \gtrdot)$ is true iff the  sentence $\phi(\textbf{x})/g$ is true for  a given  assignment $g$, that is, for any Herbrand interpretation $v$ it must hold the following truth constraint:
\begin{equation}\label{eq:truthMV}
v(T(\lessdot\phi(\textbf{x})/g \gtrdot)) = v^*(\phi(\textbf{x})/g)
\end{equation}
 We denote by $\I_{MV} \subseteq X^{\L_0}$ the set of all many-valued valuations
 of $\L_{in}$ that have fixed (invariant) interpretation for each built-in predicate.\footnote{As, for example, the identity predicate $p_1^2 \in P$
 or nullary predicate letters $\{p_a~|~ a \in X\}$ for which it must be satisfied that for any
 $v^* \in \I_{MV}, v^*(p_a) = a$,  $v^*(p_1^2 (u_1,u_2)) = t$ iff $(u_1,u_2,t) \in  R_{=}$, and $v^*(p_2^2 (u_1,u_2)) = t$ iff $(u_1,u_2,t) \in
 R_{=_{in}}$,  and for any assignment $g$,
 $v^*(p_1^1 (t_i)) = v^*(T(t_i)) =_{def}  v^*(\phi/g)$ where the term $t_i = \lessdot \phi/g\gtrdot$.\\
 The difference between  $\I_{MV}$ and the total set of Herbrand interpretations $X^H$ is caused by the presence of the built-in predicates and constraints in (\ref{eq:truthMV}).}
\end{definition}
We recall that the set-based (for infinite sets as well) of the operators $\bigwedge$ and $\bigvee$ is well defined because our many-valued logics are based on the complete (and distributive) lattices $(X, \leq)$, which satisfy these requirements.

Clearly, the many-valued quantifiers defined in definition above
does not satisfy the standard FOL property that $\forall = ~\neg
\exists \neg$, but in the case of the 2-valued logic when $X =
\textbf{2}$, then this requirement is satisfied by Definition
\ref{def:MV-algebra} and points 3 and 4 corresponds to the standard semantics of quantifiers of the FOL.

The main difference with standard FOL syntax is that here we can use
abstracted terms obtained from logic formulae, for example, "x believes that $\phi$" is given by formula $p_i^2(x,\lessdot \phi \gtrdot)$ ( where $p_i^2$ is binary "believe" predicate).

Let $\phi(\textbf{x})$ be any well-formed virtual predicate in $\L_{in}$,
then $\lessdot \phi(\textbf{x})\gtrdot_{\alpha}^{\beta}/g$  with the set of hidden variables $\alpha = (x_1,...,x_m)$, $m \geq 1$, is the  ground term (the assignment $g$ is applied only to free visible variables in $\beta = \overline{\textbf{x}}-\alpha$ whose semantics correlate is an intensional entity (concept) of degree $m$. If $m =0$, the intensional correlate of this singular ground term
is the proposition (sentence) "that $\phi$"; if $m =1$, the intensional
correlate is the property of "being something $x_1$ such that $\phi$";
if $m > 1$, then the intensional correlate is the concept
"the relation among $x_1,...,x_m$ such that $\phi$".

Certain complex nominative expressions (namely, gerundive and
infinitive phrases) are best represented as singular terms of the
sort provided by our generalized bracket notation $\lessdot
\phi(\textbf{x})\gtrdot_{x_1,...,x_m}^{\beta}$, where $m \geq 1$. This MV-intensional logic
$~\L_{in}$ differs from two-valued intensional FOL  in heaving these singular terms
$\lessdot \phi(\textbf{x})\gtrdot_{x_1,...,x_m}^{\beta}$ where the virtual predicate $\phi(\textbf{x})$ is \emph{many-valued}.

 The $IFOL_B$ is a \emph{many-sorted }logic. We introduce the sorts in order to  assign to each variable $x_i$  a sort $s_i \in \S$. Set of sorts $\S$ is composed by elements of a domain $\D$ as well (individual \emph{atomic} sorts which are the names of concepts in $\D$ are usually denoted by $s_i \in \S$). We denote by $\L$ and $\T$ the set of all formulae and terms in his logic $\L_{in}$, respectively.

\section{Many-valued  Algebras of Concepts and their Extensions \label{sec:MVFOLalg}}
 We assume, as in \cite{Majk22},  that a concept algebra of many-valued logic $IFOL_B$ has a non empty domain (it is different from the domain of 2-valued standard IFOL) $~\D = D_{0} + D_I$, (here $+$ is a disjoint union) where a subdomain $D_{0}$ is made of  particulars or individuals (we denote by $\circledR$ the non-meaning individuals, for interpretation of language entities that are no meaningful, as "Unicorn" for example) with $X \subseteq D_0$.

The rest $D_I = D_1 + D_2 ...+ D_n ...$ is made of
 universals (concepts): $D_1$ for many-valued logic sentences, as described in Section 2 of \cite{Majk24ar}, called \verb"L-concepts" (their extension corresponds to some logic value),   and  $D_n, n \geq 2,$ for
 \verb"concepts" (their m-extension is an n-ary relation); we consider the property (for an unary predicate) as a  concept in $D_2$. The  concepts in $\D_I$ are denoted by $u,v,...$, while the  values (individuals) in $D_0$ by $a,b,...$\\
 \textbf{Remark}: Notice that all concepts in $D_i$, $i\geq 2$ corresponds to the \emph{ontologically encapsuled} virtual predicates (as in the case of the ontological encapsulation  \cite{Majk22} of many-valued atoms $p_i^{n-1}(x_1,...,x_{n-1})$ by extending them with another attribute to obtain extended 2-valued "flattened" atom  $p_F(x_1,...,x_{n-1},y)$ where domain of $y$ is the set of logic-values in $X$).
 In this way, the last $(k+1)$-th attribute of the extension of any  concept in $D_{k}$, $k\geq 2$, will have the assigned logic value to the original  many-valued $k$-ary concept of the many-valued predicate logic $\L_{mv}$ for a given valuation $v$.
 \\$\square$\\
 \textbf{Basic concepts derived from logic propositions and atoms}:\\
 The intensional interpretation $I$ maps each nullary predicate symbol (proposition symbol) into $D_1$, -and for each built-in proposition symbol of $IFOL_B$ syntax (a nullary predicate) $p_a$, for $a \in X$, we have corresponding  built-in truth-concept
 \begin{equation}  \label{eq:truth-conc}
  u_a = I(p_a)\in D_1
\end{equation}
 The language constants (nullary functional symbols in $P$) are mapped in
concepts in $\D$. Notice that for any non-meaningful language constant (as "Unicorn" for example) we have that $I(c) = \circledR \in D_0$.

 Consequently, each atom $p_i^k(t_1,...,t_k)$, $p_i^k \in P$, is mapped into
the concept $u \in D_{m+1}$, where $m$ is a number of the free
variables of the virtual predicate obtained from this atom.
Consequently, each ground atom $p_i^k(t_1,...,t_k)$ is mapped into
$D_1$, and if there is any $t_i$, $1 \leq i \leq k$, such that
$I(t_i) = \circledR$, then $I(p_i^k(t_1,...,t_k)) = u_{\bot}$, where
$\bot \in X$ such that $\neg \bot = \bot$ is the
logic truth-value 'unknown'. In this way we guarantee that unmeaningful sentences as
$blu(Unicorn)$ will always have the 'unknown' logic truth-value. This
intensional interpretation can be given also to all contradictory
formulae (with truth-value $\top\in X$) that can not be nor true nor false, as the Liars paradoxes.

 We have that $I(p_1^2(x,y)) = I(x = y) = Id \in D_3$, while $I(p_1^2(x,y)/g) = I((x = y)/g) = pred_3(g(x),g(y),t,Id) \in D_1$.
Analogously, we have that \\$I(p_2^2(x,y)) = I(x =_{in} y) = Eq \in
D_3$, while $I(p_2^2(x,y)/g) = I((x =_{in} y)/g) =
pred_3(g(x),g(y),t,Eq) \in D_1$.  The atom  $T(x)$ of the truth predicate $T$ is mapped into self-reference concept, i.e. $I(T(x)) = I(p_1^1(x)) = u_T \in D_2$.
\\$\square$\\
Let us define the extensions of the PRP concepts:
 \begin{definition} \label{def:m-sortedExtensions} \textsc{m-sorted extensions}:\\
 We define  the set of all m-extensions in the many-sorted framework of universals in $D_I$,
 \begin{equation} \label{eq:dueM5}
 \mathfrak{Rm} = \widetilde{X}\bigcup\{ R \in
~\bigcup_{n \geq 1}\P(\D^n \times
(X\backslash\{\bot\}))~|~(u_1,..., u_{n}, a), ~(u_1,..., u_{n}, b) \in R ~~implies ~~b = a \}
\end{equation}
 so that each $(n+1)$-ary relation is a graph of a function, and by  $~\mathfrak{Rm}_k, ~k \geq 1$, we will denote the subset of all k-ary relations in $\mathfrak{Rm}$, and by $\emptyset$ each empty relation in $\mathfrak{Rm}$,
with the set of unary relations in $\widetilde{X}$,
\begin{equation} \label{eq:dueM6}
\widetilde{X} =
\{\{ a\}~|~ a \in X ~~and ~~a \neq \bot \} \bigcup \{\emptyset\} =  \{ \emptyset, \{ f\}, \{ \top\},\{ t\}\}
\end{equation}
where $\{ a\}$ denotes unary relation with unique tuple equal to the truth-value $a `in X$ and $\emptyset$ unary empty relation,
and hence, $\emptyset\in \widetilde{X}\subset \P(X)$.  So, $(\widetilde{X},\preceq)$ is complete lattice with total ordering $\emptyset \preceq \{f\}\preceq \{\top\}\preceq \{t\}$, that is, with the  bottom element and   $\{a\} \preceq \{b\}$ iff $a\leq b$ (w.r.t the truth-ordering in $X$).
\end{definition}
Note that $\mathfrak{Rm}$ contains only \emph{known} extensions of the many-valued concepts, and this explains also why we defined for $\bot \in X$ the empty relation $\emptyset$.

 We introduce for the concepts in $D_I$ an \emph{extensional interpretation}  $h$ which assigns the \emph{m-extension} to each intensional concept in $\D$,
and can be considered as an \emph{interpretation of concepts} in
$\D$. Thus, each concept in $D_I$ represents a set of tuples in $\D$,
and can be also an element of the extension of another concept and
of itself also.  Each extensional interpretation $h$ assigns to the intensional elements of $\D$
an appropriate extension: in the case of particulars $u \in  D_{0}$, $h_0(u) \in D_0$, such that for each logic value $a\in X\subset D_0$, $h_0(a) = a$.  Thus, we have the particular's mapping $h_0:D_0\rightarrow D_0$ and more generally  (here $'+'$ is considered as disjoint union),
 \begin{equation}\label{eq:dueM6}
 h =   \sum_{i\in \mathbb{N}}h_i:\D \longrightarrow   D_0+\sum_{i\geq 1}\mathfrak{Rm}_i
\end{equation}
 where $h_1:D_1 \rightarrow \widetilde{X}$ assigns to each  L-concept $u \in D_1$  (of a sentence with truth-value $a\in X$), a
relation composed by the single tuple $h_1(u) = \{a\}$ if $a \neq \bot$),  $~\emptyset$ otherwise, and $h_i:D_i\rightarrow \mathfrak{Rm}_i$, for $i\geq 2$, that assigns a m-extension to non-sentence concepts.

 Set of sorts $\S$ is composed by elements of a domain $\D$ as well (individual \emph{atomic} sorts which are the names of concepts in $\D$ are usually denoted by $s_i \in \S$). For example $X \subset D_0$ is a many-valued-logic sort, $\textbf{2} \subseteq X$ is a classic-logic sort,
  $[0,1] \subset D_0$ is
 closed-interval-of-reals sort, $\mathbb{N} = \{1,2,3,..\} \subset D_0$ is a sort of positive integers, etc..
 These sorts  \cite{Majk24ar} are used for sorted variables in this many-sorted
 predicate logics so that the assigned values for each sorted variable must belong to its sort.\\
  We introduce the sorts in order to  assign to each variable $x_i$  a sort $s_i \in \S$.  An assignment $g:\V \rightarrow \D$ for variables in $\V$ is applied only to free variables in terms and formulae. For each sorted variable $x_i \in \V$ an assignment $g$ must satisfy the auxiliary condition $g(x_i) \in h(s_i)$ that this value must be one of the sort's extension. We denote by $~t_1/g~$ (or $\phi/g$) the ground term (or formula), without free variables, obtained by assignment $g$ from a
term $t_1$ (or a formula $\phi$). \\
 We use the following formal syntax structure for the concepts defined in \cite{Majk24ar}:
 \begin{definition} \label{def:var-sort}
  We assume the UNA (Unique Name Assumption) for all concepts in the PRP domain $\D$, and hence that a \emph{sort} is identified by the name (in bold letters, defined as a natural language phrase)  of corresponding concept.\\
     The syntax of a k-ary  concept  $u \in D_k$, $k\geq 1$,  is composed by its name  followed by $k\geq 1$  list of its sorts $s_i$, which are the concepts in $D_1+D_2+...+D_n $ or the special sort $'nested ~
 sentence'$ (which is not an element of the  domain $\D$), representing its properties, i.e., a concept $u\in D_k$ has a syntax form $\textbf{phrase}:s_1,...,s_k$.\\
So, if this concept $u$ is a predicate-concept (which defines a  k-ary predicate letter $p_i^k \in P$ of the FOL), there is the following relationship between the atoms of the predicates with terms $t_i$ (in this definition of predicate-concept we do not use for $t_i$  a constant (a nullary function) but a variable, in order to obtain a more general concept corresponding to this predicate letter $p_i^k \in P$),
  and corresponding predicate-concepts with the list of \emph{sorts}  $s_i$ in $\S \subset D_1+D_2+...+D_n+ \{nested ~
 sentence\}$,
 \begin{equation} \label{eq:P-Ccorrespondence}
  p_i^k(t_1,...,t_k) ~~~~~\Leftrightarrow~~~~~ \textbf{phrase}:s_1,...,s_k \in D_k
  \end{equation}
  such that if a term $t_i$ is a free variable then $s_i$ is the sort of this variable,  if $t_i$ is a functional term then $s_i$ is the sort of term's outermost functional symbol, if $t_i$ is an abstracted term then $s_i$ is the $'nested ~
 sentence'$ sort\footnote{Such abstracted term are inserted in the predicate with this variable during the parsing of a complex natural language sentences containing the nested subsentences. So, the sort $'nested ~ sentence'$, which is not an element of a domain $\D$, has the empty extension in domain $\D$ differently from all other sorts.}.\\
  Thus, we obtain the following mapping for the sorts of variables:
 \begin{equation} \label{eq:var-sort}
 \digamma:\V \rightarrow \S
 \end{equation}
   Let us define how this mapping apply to the predicate-based IFOL syntax:\\
 1. For each variable $x \in \V$, to which we assign a sort $s \in S$ (different from 'nested sentence'), denoted by $x:s$, we have that  $\digamma(x) = s$ with non-empty extension $\|s\|$.\\
 2. If the term $t_i$ is an abstracted term, then $\digamma(t_i) = 'nested ~sentence' \notin \D$.\\
 3.  For the logic constant $c$ (a nullary functional symbol) the $\digamma(t_i)$ is its defined sort.
 Otherwise,  term’s sort is the sort declared as return sort of the term’s outermost function symbol, that is, for k-ary functional symbol $f^k_i$ with $k\geq 1$ the term  $t = f_i^k(t_1,...,t_k)$ which represents function\\
 $f_i:\|\digamma(t_i)\|\times...\times \|\digamma(t_i)\| \rightarrow \|s\|$\\
 where $s \in \S$ is the sort of this functional term, that is, $\digamma(t) = s$. From the fact that in IFOL the functional symbols are represented as particular predicates, we obtain the sort of $f^k_i$ as

  $\digamma(f^k_i) = \digamma(t_1)\times...\times\digamma(t_k)\times s$.\\
 4. For each predicate symbol $p^k_i \in P\backslash F$, $k\geq 1$, corresponding to the predicate-concept $phrase:s_1,...,s_k$, its sort is

 $\digamma(p^k_i) = s_1\times...\times s_k$\\
5. For each virtual predicate $\phi_i$ (an open FOL formulae with the non empty tuple of free variables $\textbf{x}= (x_1,...x_k)$, $k\geq 1$), corresponding to composed concept by the operators in $\A_{int}$ in \cite{Majk22}, its sort is

 $\digamma(\phi_i) = \digamma(x_1)\times...\times\digamma(x_k)$.
 \end{definition}
 For any two sorts (concepts), $s_1,s_2 \in \S$, with IS-A relationship $s_1 \sqsubseteq s_2$ we say that $s_2$ is a \emph{supersort} (superconcept) of $s_1$, and $s_1$ is a \emph{subsort} (subconcept) of $s_2$.
\begin{definition} \label{def:built-in} \textsc{Built-in concepts}:
Let $\E_{in}$ be a fixed subset of extensionalization  functions (in what follows we will consider the set of all well-defined extensionalization  functions for the many-valued Belnap's intensional first-order logic $IFOL_B$, specified by   Definitions \ref{def:syntax} and \ref{def:MVintensemant}).

We say that a given concept $u \in D_I$ is a \emph{built-in concept} if its extension is invariant w.r.t. extensionalization  functions. It is the case of elementary datatypes (sorts)  as "\textbf{natural numbers}:~s", "\textbf{real-numbers}:~s", "\textbf{truth values}:~s",that are concepts in $D_2$ such that for $\mathbb{N}\subset D_0$ and $X \subset D_0$,

$h_2(\textbf{natural numbers}:~s) = \{(n,t)~|~n \in \mathbb{N}\}$,

$h_2(\textbf{truth values}:~s) = \{(a,t)~|~a \in X\} $.\\
 We define a number of another most used built-in concepts:
\begin{enumerate}
  \item We introduce the following
manyvaluedness assumption for intensional elements in $\D$ and their extensionalization  functions: for any algebraic truth value $a \in X$ we introduced by (\ref{eq:truth-conc}) a built-in truth-concept $u_a \in D_1$ such that if $a \neq \bot$

 $~~\forall h\in \E_{in} (h(u_a) = \{a\})~~$
 and  $~~\forall h \in \E_{in}(h(u_\bot) = \emptyset)$.
  \item The corresponding concept to the binary identity predicate $=$ used in FOL
 is denoted by the symbol $Id \in D_3$.  We have that for this  "concept of identity" $Id$, its m-extension is constant and 2-valued in every extensionalization function $h \in \E_{in}$, that is, $~~h(Id) = R_{=}$ where
\begin{equation}
  R_{=} =
 \left\{
    \begin{array}{ll}
    \{ (u,v,t)~|~u,v\in D_k, k \geq 0 ~~such ~that~~  u = v ~~ in ~FOL \}
     & \hbox{,}\\
       \{ (u,v,f)~|~u,v\in D_k, k \geq 0 ~~such ~that~~  u \neq v ~~ in ~FOL \} & \hbox{.}
    \end{array}
  \right.
  \end{equation}
  \end{enumerate}
\end{definition}
Another important concept that has a specific semantics, but
\emph{is not} built-in concepts is the \emph{self-reference}
truth-concept $u_T\in D_2$, such that for any particular extensionalization  function $h$ we have that $h(u_T) = \{(u,a)~|~u \in D_1,
\emptyset \neq h(u) =\{a\}\}$. This concept represents the truth of
all logic-concepts in $D_1$, in the way that for any logic-concept $u \in D_1$ we have that $~h(u) =\{a\}~$ iff $~(u,a) \in h(u_T)$.
\\
 We define the following partial ordering $\preceq$ for
the m-extensions in
$\mathfrak{Rm}$ (note that for a given k-ary relation with $k\geq 2$ and $1\leq i\leq k$, the $\pi_i(R)$ is the i-th projection of $R$ while $\pi_{-1}(R)$ a relation obtained from $R$ by eliminating i-th column of it):
\begin{definition} \label{def:POrder} \textsc{Extensional partial order in} $\mathfrak{Rm}$:\\
We extend the ordering of unary relations in $\widetilde{X }\subset\mathfrak{Rm}$ by:
for any two nonempty m-relations $R_1, R_2 \in \mathfrak{Rm}$ with arity $k_1 = ar(R_1), k_2 = ar(R_2)$ and $m = \max(k_1,k_2)\geq 2$, $\emptyset \preceq R_i$ for $i =1,2$, and

$~~~R_1\preceq R_2~~$ iff $~~$ \\for each $(u_1,...,u_{m-1},a) \in
Ex(R_1,m)$,  $\exists(u_1,...,u_{m-1},b) \in Ex(R_2,m)$
with $a \leq b$,\\
where this expansion-mapping $Ex:\mathfrak{Rm}\rightarrow
\mathfrak{Rm}$ is defined as follows for any nonempty $R \in \mathfrak{Rm}$:
\begin{equation} \label{eq:dueM7}
Ex(R,m) =
 \left\{
    \begin{array}{ll}
      \D^{m-k}\times
\{a\}, ~~for ~~R = \{a\}, ~~a \in X\backslash\{\bot\} & \hbox{if $~~m > k=1$}\\
  \bigcup \{\{(u_1,...,u_k)\}\times \D^{m-k}\times
\{a\}~|~(u_1,...,u_k,a) \in R\}, & \hbox{if $~~m > k> 1$}\\
      R, & \hbox{otherwise}
    \end{array}
  \right.
\end{equation}
We denote by $R_1 \simeq R_2$ iff $R_1\preceq R_2$ and $R_2\preceq
R_1$, the equivalence relation between m-extensions.
We define the completion of  $R \in \mathfrak{Rm}$  with $~i = ar(R)\geq 1$ by the mapping $Com:\mathfrak{Rm}\rightarrow \mathfrak{Rm}$,
\begin{equation} \label{eq:dueM8}
Com(R) =
 \left\{
    \begin{array}{ll}
    D^{i-1}\times \{\bot\}, & \hbox{if $~i \geq 2$ and R empty}\\
  R\bigcup \{(v_1,..,v_{i-1},\bot) ~|~ (v_1,..,v_{i-1}) \notin
\pi_{-i}(R)\}, & \hbox{if $~i \geq 2$ and R nonempty} \\
      \{\bot\}, & \hbox{if $R  = \emptyset$ and $~ i =1)$}\\
      R, & \hbox{otherwise }
    \end{array}
  \right.
\end{equation}
 We define the inclusion  mapping $Inc:\mathfrak{Rm}\times \mathfrak{Rm}\rightarrow
\mathfrak{Rm}$ by:\\
 $~~~Inc(R_1,R_2) = \{t\}~$ if $~ar(R_1) = ar(R_2)$ and there is a relation $R_2'$ obtained by a permutation of columns in $R_2$ such that $R_1
\subseteq R_2'$; $~\{f\}$ otherwise.
\end{definition}
We chose $\emptyset, \{t\} \in  \widetilde{X} \subset \mathfrak{Rm}$ to be the representative elements of the bottom and top equivalence classes in $\mathfrak{Rm}$ relatively (determined by $\simeq$), so that $(\forall R \in \mathfrak{Rm})( \emptyset \preceq R \preceq \{t\})$. Consequently, the \emph{extensional} ordering $\preceq$ in $\mathfrak{Rm}$ extends the ordering of the lattice  $(\widetilde{X},\preceq)$.

In what follows for any two relations $R_1,R_2 \in \mathfrak{Rm}$, for each set of couples of indexes $S \in
\P(\mathbb{N}^2)$,  in the algebraic expression $R_1\bowtie_S R_2$, the binary operation $\bowtie_S$ is the natural
join from relational database algebra if $S$ is a non empty set of
pairs of joined columns of respective relations (where the first
argument is the column index of the relation $R_1$ while the second
argument is the column index of the joined column of the relation
$R_2$); \verb"otherwise" it is equal to the cartesian product
$\times$. We can define an m-extensional algebra over the poset
$(\mathfrak{Rm}, \preceq)$ as follows:
\begin{definition} \label{def:m-extensions}
Let us define the m-extensional relational algebra for   Belnap's truth-values algebra $\A_{B} =(X, \leq, \neg,\wedge,\vee,\Rightarrow,\Leftrightarrow)$ of Definition \ref{def:mv-algebra} by,\\
$\A_{\mathfrak{Rm}} = ((\mathfrak{Rm}, ~\preceq, ~Com, ~Inc, ~\emptyset, ~\{t\}),
~R_=,  ~\oslash, ~\{\otimes_S, \oplus_S,\ominus_S,\}_{ S \in
\P(\mathbb{N}^2)}, \circledast, ~\{\boxplus_i, \boxtimes_i\}_{ i \in \mathbb{N}})$, \\where  $R_=$ is the binary
relation for extensionally equal elements and $\circledast =   \otimes(Inc(\_),Inc(\_)^{-1}):\mathfrak{Rm}^2
 \rightarrow \mathfrak{Rm}$ derived binary operator where $\otimes$ is the binary operator $\otimes_S$ for empty set $S$.
We will use '$=$' for the extensional identity for relations in $\mathfrak{Rm}$.
The unary operators $\oslash, \boxplus_i, \boxtimes_i: \mathfrak{Rm}
 \rightarrow \mathfrak{Rm}$, and binary operators
$\otimes_S, \oplus_S, \ominus_S:\mathfrak{Rm}^2
 \rightarrow \mathfrak{Rm}$,
are defined as follows for any $R,R_1,R_2 \in \mathfrak{Rm}$ with arity greater than zero:
 \begin{enumerate}
   \item For negation $\oslash$:\\
   1.1 $~~~\oslash(R) =  \{\neg a  ~|~ \{a \} = R, ~\neg a \neq \bot \} $, $~~$ for $ar(R) =1$.\\
1.2 $~~~\oslash(R) = \{(v_1,..,v_{i-1},\neg a ) ~|~
(v_1,..,v_{i-1},a)
\in Com(R), ~\neg a \neq \bot \}$,\\ $~~$ for $~i = ar(R) \geq 2$.\\
1.3 $~~~\oslash(\emptyset) = \emptyset$.
   \item For any pair of binary operations $(\odot,\circ) \in \{(\otimes_S,\wedge),
(\oplus_S,\vee), (\ominus_S, \Rightarrow)\}$:\\
2.1 $~~~R \otimes_S \emptyset  = \emptyset \otimes_S R = \emptyset$; $~~~R \oplus_S \emptyset  = \emptyset \oplus_S R = R$; $~~~\emptyset \ominus_S \emptyset  = \{t\}$;\\
$~~~\emptyset \ominus_S \{a\} = \{\bot \Rightarrow a\}$ if $\bot \Rightarrow a \neq \bot$, $\emptyset$ otherwise;\\
$~~~\{a\} \ominus_S \emptyset = \{a \Rightarrow \bot\}$ if $a \Rightarrow \bot \neq \bot$, $\emptyset$ otherwise. And for $i =ar(R)\geq 2$,\\
$~~~\emptyset \ominus_S R = \{(v_1,...,v_{i-1},\bot \Rightarrow a)|(v_1,...,v_{i-1}, a)\in R$ and $\bot \Rightarrow a \neq \bot \}$;\\
$~~~R \ominus_S \emptyset = \{(v_1,...,v_{i-1},a \Rightarrow \bot)|(v_1,...,v_{i-1}, a)\in R$ and $a \Rightarrow \bot \neq \bot \}$;\\
 2.2 $~~~R_1 \odot R_2 = \{a \circ b ~|~ \{a\} = Com(R_1),
\{b\} = Com(R_2)$ and  $a \circ b \neq \bot\} $, \\$~~$ for $ar(R_1) = ar(R_2) = 1$ and $S$ is the empty set. \\
2.3 $~~~R_1 \odot R_2 = \{(v_1,..,v_{i-1},a \circ b) ~|~
(v_1,..,v_{i-1},a) \in Com(R_1), \{b\}\in Com(R_2)$ and $a \circ
b\neq \bot\} $,\\ $~~$ for $ar(R_1) = i \geq 2, ar(R_2) = 1$ and $S$ is the empty set.\\
2.4 $~~~R_1 \odot R_2 = \{(v_1,..,v_{i-1},a \circ b) ~|~\{a\}\in
Com(R_1), (v_1,..,v_{i-1},b) \in Com(R_2)$ and  $a \circ b \neq
\bot\} $,\\ $~~$ for $ar(R_1) = 1, ar(R_2) = i \geq 2$ and $S$ is the empty set.\\
 2.5 Otherwise: Let $i = ar(R_1) \geq 2, ar(R_2)   \geq 2$, and we use the binary operator $~\bowtie_{S}:\mathfrak{Rm} \times \mathfrak{Rm} \rightarrow \mathfrak{Rm}$,
 such that for any two relations $R_1, R_2 \in  \mathfrak{Rm}~$, the
 $~R_1 \bowtie_{S} R_2$ is equal to the relation obtained by natural join
 of these two relations $~$ \verb"if" $S$ is a non empty
set of pairs of joined columns of respective relations (where the
first argument is the column index of the relation $R_1$ while the
second argument is the column index of the joined column of the
relation $R_2$); \verb"otherwise" it is equal to the cartesian
product $R_1\times R_2$.  Then,\\
$~R_1 \odot R_2 = \{(v_1,..,v_{k},a\circ b) ~|~ ((v_1,..,v_{i-1},a,v_{i },...,v_{k},b) \in Com(R_1)\bowtie_S Com(R_2)$ with $(v_1,..,v_{i-1},a) \in R_1$  and $(v_{i },...,v_{k},b)$ is a (also empty) sub-tuple of $R_2$ without elements in $\pi_2(S)$)  and  $a \circ b \neq \bot\}$.
   \item For  quantifiers $\exists_i$ and $\forall_i$, existential $~\boxplus_i$ and universal $~\boxtimes_i$, (here '$\bigwedge$' and ;$\bigvee$; are the meet and join operators in lattice of truth values $X$, respectively):\\
   3.1 $~~$ for $~ar(R) = 1$, $~~~\boxplus_i(R) = \boxtimes_i(R) = R$.\\
3.2 $~~$ for $ar(R) = 2$,\\
$~~~\boxplus_i(R) = \{b ~|~ $if $~b = \bigvee \{a ~|~(v,a) \in
Com(R)\} \neq \bot\}$ if $~i = 1$; $~R$ otherwise,\\
$~~~\boxtimes_i(R) = \{b ~|~ $if $~b = \bigwedge \{a ~|~(v,a) \in
Com(R)\} \neq \bot\}~$ if $~i = 1$; $~R$ otherwise.\\
 3.3 $~~$ for $ar(R) \geq 3$,\\
 $~~~\boxplus_i(R) = \{(v_1,...,v_{i-1},v_{i+1},...,v_k,b) ~|~
$if $~b = \bigvee \{a ~|~(v_1,...,v_k,a) \in Com(R)\} \neq \bot\}~$ if $~1
\leq i \leq k$; $~R$ otherwise,\\
  $~~~\boxtimes_i(R) = \{(v_1,...,v_{i-1},v_{i+1},...,v_k,b) ~|~
$if $~b = \bigwedge \{a ~|~(v_1,...,v_k,a) \in Com(R)\} \neq \bot\}~$ if
$~1 \leq i \leq k$; $~R$ otherwise.
 \end{enumerate}
\end{definition}
Notice that definition of operators $\boxplus_i, \boxtimes_i$ are
valid because $X$ is \emph{complete} lattice.

Moreover, we define the many-valued-ISA relationship for concepts in
$u, v \in D_I$ by $~u \sqsubseteq v~$ iff  $~$ for all $h$,
$(h(u)\preceq h(v))$, which is a generalization of the ISA
relationship in DL (Description Logic). The strong intensional equality, denoted by
$u \circeq v$ (i.e., $(u,v) \in ~\circeq$),  holds if $(u \sqsubseteq
v)$ and $(v \sqsubseteq u)$.

So, we introduce the following  algebra for concepts for this IFOL with Belnap's bilattice of logic truth-values,  and able to support the unknown and inconsistent knowledge for AGI robots:
\begin{definition}  \label{def:Int-algebra} Algebra for concepts of $IFOL_B$ is a structure\\ $~\mathcal{A}\mathfrak{B}_{int} =
\langle(\D,   \sqsubseteq,    \{pred_i\}_{i \in \mathbb{N}}, u_\bot, u_t), Id,  \widetilde{\neg},
\{\sqcap_{S}, \sqcup_{S}, \widetilde{\Rightarrow}_S \}_{ S \in
\P(\mathbb{N}^2)},  \leftrightarrow, \{\widetilde{\exists}_i,  \widetilde{\forall}_i\}_{i \in \mathbb{N}}\rangle$,\\
 with unary negation, exists, and forall   operations  $~~\widetilde{\neg}, \widetilde{\exists}_i, \widetilde{\forall}_i:D_I\rightarrow D_I$, and with binary operation  for the intersection,
union, implication and strong equivalence of concepts $\sqcap_{S},\sqcup_{S},
\Rightarrow_S, \leftrightarrow:D_I\times D_I \rightarrow D_I$,  and operations $~ pred_i:\D^{i-1}\times X\times D_i \rightarrow D_1$.\\
The semantics of these intensional algebraic operations for any
given extensionalization  function $h:D_I \rightarrow \mathfrak{Rm}$ is
given as follows:
\begin{enumerate}
\item $~~~h(u_\bot) = \emptyset$, $h(u_t) = \{t\}$, $h(Id) = R_=$, where $R_=$ is the binary relation of extensionaly equal elements (as in standard FOL).
     \item $~~~h(u \leftrightarrow v) =  \circledast(h(u),h(v)) =\\ = \otimes(Inc(h(u),h(v)),Inc(h(v),h(u)))\in \{\{f\}, \{t\}\}$.
  \item $~~~h(pred_k(v_1,..,v_k,u)) = Inc(\{(v_1,..,v_{k})\},
Com(h(u))) \in \{\{f\}, \{t\}\}$,\\ $~$ where $u \in D_k$.
That is, the predication that the tuple $(v_1,..,v_{k})$, with $v_k \in X$ (we used completion operation $'Com'$ because we can have $v_k =\bot$ as well), is an element of the extension of the $k$-ary concept $u$.
  \item $~~~h(\widetilde{\neg} u) =  \oslash (h(u))$.
  \item For any pair $(\odot,\circ) \in \{(\sqcap_S,\otimes_S),
(\sqcup_S,\oplus_S), (\widetilde{\Rightarrow}_S, \ominus_S)\}$,
 $~~~h(u \odot v) = h(u) \circ h(v)$.
   \item $~~~h(\widetilde{\exists}_iu) = \boxplus_i(h(u)), ~~~h(\widetilde{\forall}_iu) = \boxtimes_i(h(u))$ for quantifiers.
\end{enumerate}
\end{definition}
Notice that  if for all $h$, $h(u \leftrightarrow v) = \{t\}$, then $u \equiv v$, that is, these two concepts are strongly equivalent. The relationships between the \emph{identity} of $u $ and $v$ (i.e.,
when $h(pred_3(u,v,t,Id)) = \{t\}$) and (strong) \emph{equivalence}
$u \equiv v$  is the standard relationship: if $u$ is identical to
$v$ then $u \equiv v$ but not vice versa: let us consider two
different concepts $u, v \in D_2$ for "has been sold" and "has been bought" respectively, we have that $u$ is not identical to $v$ (that is, $h(pred_3(u,v,t,Id)) = \{f\}$), while $u \equiv v$ from the fact that for all $h$, $h(u) = h(v)$.\\
\textbf{Remark}: By these operators it is possible also to make meta-sentences
about sentences, that is,  if $a$ is a truth-value for the L-concept (of same sentence) $u \in D_1$ ( i.e., if $h(u) = \{a\}$) then the metalogic sentence "\emph{it is true that $u$ has a logic truth-value} $a$" corresponds to the fact that $h(pred_1(a,u)) = \{t\}$. If $h(pred_1(\bot,u)) = \{t\}$ this means that truth-value of this L-comcept (sentence) is unknown. Thus, the robot is able for any sentence to deduce if it is unknown, or if it is known fact which truth-value $a \in \{f, \top, t\}$ it has in each instance of time (during time the robot's experience changes so, for example,  the unknown state can become false, inconsistent (possible) or true fact. Thus, it holds that for the \emph{self-reference truth-concept} $u_T \in D_2$,
$h(pred_2(u,a,u_T)) = h(pred_1(a,u))$.  So, for actual extensionalization function $h$, each AGI robot is able to know the truth-value $a \in X$ of each of its L-concepts (corresponding to its generated sentences) and to communicate it by using natural language to humans or other AGI robots.
\\$\square$\\
The following important homomorphism between intensional and extensional algebras,  \index{homomorphism of algebras} provided by Definition \ref{def:Int-algebra} and Definition \ref{def:m-extensions}, there exists:
\begin{coro}  \label{coro:intens} The semantics of the algebra for concepts in Definition \ref{def:Int-algebra} extends an extensionalization function $h$ given by (\ref{eq:dueM6}) into the homomorphism from intensional into the m-extensional algebra (in Definition \ref{def:m-extensions}):
\begin{equation} \label{eq:int-ExtHomomorph}
h:\mathcal{A}\mathfrak{B}_{int}\rightarrow \A_{\mathfrak{Rm}}
\end{equation}
We denote by $\E$ the set of all homomorphisms between these two
algebras, represented by the following set-mapping:
\begin{equation} \label{eq:h-homomorphisms}
~\mathcal{A}\mathfrak{B}_{int}\Longrightarrow_{h\in \E} \A_{\mathfrak{Rm}}
\end{equation}
\end{coro}
\textbf{Proof:} Directly from Definition \ref{def:Int-algebra}: it
preserves the ordering, i.e. if $u \sqsubseteq v$ then $h(u)\preceq
h(v)$, with $h(u_\bot) = \emptyset, h(u_t) = \{t\}$ for bottom/top
elements, and $h(Id) = R_{=}$. \\
For algebraic operations we have that \\
$h(\leftrightarrow) = \circledast$, $~h(pred_i(v_1,..,v_i,\_)) = Inc(\{(v_1,..,v_{i})\}, Com(\_))$, $h(\widetilde{\neg}) =  \oslash$,
$h(\sqcap_S) = \otimes_S, h(\sqcup_S) =\oplus_S, h(\widetilde{\Rightarrow}_S) =
\ominus_S$,   $~h(\widetilde{\exists}_i) =
\boxplus_i, h(\widetilde{\forall}_i) = \boxtimes_i$.
\\ $\square$
\section{Interpretations of $IFOL_B$ with Compositional  Montague's Intension \label{section:intensionalMV}}

As in the case of  two-valued intensional FOL, in this many-valued case we will use the \emph{virtual predicates} provided by Definition \ref{def:virt-predicate} useful for intensional mapping into  concepts in $\D_I$, i.e., into intensional algebra $\mathcal{A}\mathfrak{B}_{int}$ provided by Definition \ref{def:Int-algebra}.

 Now we can consider the intensional semantics of $\L_{in}$, based on the translation of this logic into the intensional algebra  of concepts $\mathcal{A}\mathfrak{B}_{int}$, given by Definition \ref{def:Int-algebra}.
 We have an algebraic two-step  semantics based on the two consecutive interpretations:
  intensional interpretation of logic formulae into intensional entities (concepts) and extensional semantics of concepts given by Corollary \ref{coro:intens} in previous Section.

The intensional interpretation $I$ defined in previous sections for transformation of propositions and logic atoms into atomic concepts can be extended to all formulae of $IFOL_B$ as a following homomorphism between syntax and concept algebras:
\begin{coro} \label{coro:intensemantMV}
The intensional interpretation defines the homomorphism between free
syntax FOL language algebra $\mathcal{A}\mathfrak{B}_{FOL} = (\L, \neg, \{\wedge_{S}, \vee_{S}, \Rightarrow_S \}_{ S \in
\P(\mathbb{N}^2)}, \Leftrightarrow, \{\exists_i, \forall_i\}_{i\in \mathbb{N}})$, representing the 4-valued many-sorted intensional first-order logic  $\L_{in}$ in Definition  \ref{def:syntax}, into the intensional algebra $\mathcal{A}\mathfrak{B}_{int}$,
\begin{equation}\label{eq:IntensHomom}
I:\mathcal{A}\mathfrak{B}_{FOL}  \rightarrow \mathcal{A}\mathfrak{B}_{int}
\end{equation}
 \end{coro}
\textbf{Proof}:
 We define the following extension of the intensional
interpretation from atoms to all formulae $I:\L \rightarrow \D$
(notice that in this recursive definition we are using virtual
predicates obtained from open formulae in the set of all formulae $\L$):
\begin{enumerate}
\item The logic formula $\neg \phi(x_i,x_j,x_k,x_l,x_m)$ will be
intensionally interpreted by the concept $u_1  \in D_6$, obtained by
the algebraic expression $~\widetilde{\neg} u$ where \\$u =
I(\phi(x_i,x_j,x_k,x_l,x_m)) \in D_6$ is the concept of the
virtual predicate $\phi$. Consequently, we have that for any formula
$\phi \in \L$, $~~I(\neg \phi) = \widetilde{\neg}(I(\phi))$.
  \item The algebraic expression $\phi(x_i,x_j,x_k,x_l,x_m) \wedge_S \psi
(x_l,y_i,x_j,y_j)$ (representing the virtual predicate of logic formula $\phi(x_i,x_j,x_k,x_l,x_m) \wedge \psi
(x_l,y_i,x_j,y_j)$), with $S = \{(4,1),(2,3)\}$, will be intensionally interpreted by the concept
$u_1 \in D_8$, obtained by the algebraic expression $~ u \sqcap_{S}
v$ where $u = I(\phi(x_i,x_j,x_k,x_l,x_m)) \in D_6, v = I(\psi
(x_l,y_i,x_j,y_j))\in D_5$ are the concepts of the virtual
predicates $\phi, \psi$, relatively.
Consequently, we have that for any two formulae $\phi,\psi \in \L$
and a particular  operator  uniquely determined by tuples of free
variables in these two formulae, $I(\phi \wedge_S \psi) =
I(\phi)\sqcap_{S} I(\psi)$, and, analogously, \\$I(\phi \vee_S \psi) =
I(\phi)\sqcup_{S} I(\psi)$, $I(\phi \Rightarrow_S \psi) =
I(\phi)\widetilde{\Rightarrow}_{S} I(\psi)$.\\
If $\phi$ and $\psi$ are sentences,  or ground formulae $\phi/g$ and $\psi/g$ for a given assignments $g$ (see in next Definition \ref{def:MV-algebra} ), then $S$ is empty set $\emptyset$ and we can use the simple formulae expression $\phi \circledcirc \psi$ where $\circledcirc \in \{\wedge, \vee, \Rightarrow\}$ are binary operations of Belnap's 4-valued truth-lattice $X$.
\item For any two formulae $\phi$ and $\psi$,  \\
$I(\phi \Leftrightarrow \psi) = I(\phi) \leftrightarrow I(\psi)$.
\item The logic formula  $\psi$ equal to the formula $(\exists_i) \phi(\textbf{x})$ where   $\phi(\textbf{x})$ is a virtual predicate with $\textbf{x} = (x_1,...,x_m)$,
  and $x_i \in (x_1,...,x_m)$ is i-th variable in $\phi$ . Then, we obtain the virtual   predicate $\psi(x_1,...,x_{i-1},x_{i+1},...,x_m)$, so that $I(\psi) = I((\exists_i) \phi) = \widetilde{\exists}_i u \in D_{m}$ where $u = I(\phi) \in D_{m+1}$.\\
    For example, the logic formula $(\exists_3) \phi(x_i,x_j,x_k,x_l,x_m)$ will
be intensionally interpreted by the concept $u_1  \in D_5$, obtained
by the algebraic expression $~\exists_{3}u$ where $u =
I(\phi(x_i,x_j,x_k,x_l,x_m)) \in D_6$ is the concept of the virtual
predicate $\phi$. Thus, we have that for any formula
$(\exists_n)\phi \in \L$, the operator $\exists_{n}$ is uniquely
determined by the n-th position of the  existentially quantified variable
in the tuple of free variables in the virtual predicate $\phi$,
$~I((\exists_n)\phi) = \widetilde{\exists}_n(I(\phi))$, and, analogously,
$~I((\forall_n)\phi) = \widetilde{\forall}_n(I(\phi))$.
\end{enumerate}
$\square$\\
By using the intensional mapping $I$, we are able to extend the simple assignment to variables to all (also abstracted) terms:
\begin{definition}
 An assignment $g:\V \rightarrow \D$ forb variables in $\V$ is applied only to free variables in terms and formulae.  Such an assignment $g \in \D^{\V}$ can be recursively uniquely extended into the assignment $g^*:\T \rightarrow \D$, where $\T$ denotes the set of all terms (here $I$ is an intensional interpretation of this FOL, as explained in what follows), by :
\begin{enumerate}
  \item $g^*(t) = g(x) \in \D$ if the term $t$ is a variable $x \in\V$.
  \item $g^*(t) = I(c) \in \D$ if the term $t$ is a constant (nullary functional symbol) $c\in P$.
  \item If $t$ is an abstracted term obtained for an open formula $\phi_i$, $\lessdot \phi_i(\textbf{x}_i) \gtrdot_{\alpha_i}^{\beta_i}$,  then we must restrict the assignment to $g\in \D^{\beta_i}$ and to obtain recursive definition (when also $\phi_i(\textbf{x}_i)$ contains abstracted terms:
\begin{equation} \label{eq:assAbTerm}
  g^*(\lessdot \phi_i(\textbf{x}_i)\gtrdot_{\alpha_i}^{\beta_i}) =_{def}
    \left\{
    \begin{array}{ll}
   I(\phi_i(\textbf{x}_i))~~ \in D_{|\alpha_i|+1}, & \hbox{if  $\beta_i$ is  empty}\\
       I(\phi_i(\textbf{x}_i)[\beta_i
/g(\beta_i)])~~ \in D_{|\alpha_i|+1}, & \hbox{otherwise}
       \end{array}
  \right.
 \end{equation}
where $g(\beta) = g(\{y_1,..,y_m\}) = \{g(y_1),...,g(y_m)\}$ and $[\beta
/g(\beta)]$ is a uniform replacement of each i-th variable in the
set $\beta$ with the i-th constant in the set $g(\beta)$. Notice that $\alpha$ is the set of all free variables in the formula $\phi[\beta /g(\beta)]$.
\item  If $~t = \lessdot \phi_i\gtrdot$ is an abstracted term obtained from a sentence $\phi_i$ then \\$g^*(\lessdot \phi_i\gtrdot) = I(\phi_i) \in D_0$.\footnote{
 This case 4 is the particular case 3 when the tuple of variables $\textbf{x}_i$ is empty and hence  $\beta_i$ and $\alpha_i$ are empty sets of variables with $|\alpha_i| = 0$.}
\end{enumerate}
\end{definition}
We denote by $\L_0$ the subset of all sentences \emph{in the syntax algebra } $\mathcal{A}\mathfrak{B}_{FOL}$ of this logic $\L_{in}$, that is, $\L_0 \subset \L$.\\
  Now we are able to define formally the intensional semantics of this logic:
 \begin{definition} \label{def:MVintensemant} \textsc{Two-step \textsc{A}lgebraic \textsc{I}ntensional \textsc{S}emantics \cite{Majk11TS}:}
 The intensional semantics of the logic  $\L_{in}$ with the set of formulae $\L$ can be represented by the  mapping
\begin{equation}\label{eq:2-step}
~~ \L ~\longrightarrow_I~ \D ~\Longrightarrow_{h \in ~\E_{in} \subseteq~
\E}~ \mathfrak{Rm}
\end{equation}
where $~\longrightarrow_I~$ is a \emph{fixed intensional} interpretation (homomorphism in Corollary \ref{coro:intensemantMV}) $I:\L \rightarrow \D$ and $~\Longrightarrow_{h \in \E_{in}}~$ is \emph{the subset} of all homomorphisms $h \in \E$ in Corollary
\ref{coro:intens}, such that respect all built-in concepts in Definition \ref{def:built-in}, and  for any virtual predicate $\phi(x_1,...,x_n) \in \L$, the following 4-valued generalization of Tarski's FOL constraint is valid:
\begin{multline} \label{eq:MVtarski}
 h(I(\phi(x_1,...,x_n))) =_{def} \\\{(g(x_1),...,g(x_n),a)~|~g \in
 \D^{\V},  h(I(\phi(x_1,...,x_n)/g)) = \{a\}, a\neq \bot \}
\end{multline}
  with $h(I(\phi(x_1,...,x_n)/g)) = \emptyset~~$ iff $~~
(g(x_1),...,g(x_n))\notin \pi_{-n-1}(h(I(\phi(x_1,...,x_n))))$,\\
This set $\E_{in}$ of homomorphisms $h:\mathcal{A}\mathfrak{B}_{int} \rightarrow \A_{\mathfrak{Rm}}$ will be called as possible worlds as well. We denote by $\textbf{h}$ the current extensionalization function in a current fixed instance of time.
\end{definition}
Notice that for any $h \in \E_{in}$ and a permutation $\lambda$,
$Inc(h(\phi(x_1,...,x_m)_{\lambda}),   h(\phi(x_1,...,x_m)))\\ = Inc(h(\phi(x_1,...,x_m)),   h(\phi(x_1,...,x_m)_{\lambda})) = \{t\}$.

Notice that Belnap's 4-valued extended interpretation $v^*:\L_0 \rightarrow X$ in Definition \ref{def:MV-algebra} defined previously \emph{is not a homomorphism}, because of the point 3 (and 4 as well), where the truth of the closed formula $(\forall x_i) \phi(x_i)$ can not be obtained from the logic value of $\phi(x_i)$ from the fact that  to any formula with free variables we can not associate any logic value. Because of that, we define a new version of many-valued interpretation, denominated "MV-interpretation":
\begin{definition} \label{def:MV-interpret} \textsc{MV-interpretations}:
We define, for a valuation $v^*:\L_0\rightarrow X$ of the sentences in $\L_0 \subset \L$ of the 4-valued many-sorted intensional first-order logic  $\L_{in}$ in Definition \ref{def:syntax},  the MV-interpretation $I^*_{B}:\L_0 \rightarrow \mathfrak{Rm}$, such that for any sentence $\phi/g\in \L_0$,
 \begin{equation}\label{eq:mvR}
 I^*_{B}(\phi/g) =
 \left\{
    \begin{array}{ll}
 \{v^*(\phi/g)\} \in \widetilde{X} \subset \mathfrak{Rm}, & \hbox{if $~~v^*(\phi/g) \neq \bot$}\\
      \emptyset, & \hbox{otherwise}
    \end{array}
  \right.
 \end{equation}
 and we define also the unique extension of $I^*_{B}$ to all open formulae $\L$ in
$\L_{in}$ as well, such that for any open formula (virtual predicate) $\phi(x_1,...,x_k) \in \L$,
\begin{equation} \label{eq:openEQ}
I^*_{B}(\phi(x_1,...,x_k)) = \{(g(x_1),...,g(x_k),a)~|~g \in
\D^{\V}~~ and~~ a = v^*(\phi/g)  \neq \bot\}\in \mathfrak{Rm}
\end{equation}
We denote by $\W$ the set of all MV-interpretations derived from the set of 4-valued interpretations $v^*\in \I_{MV}$ specified in Definition \ref{def:MV-algebra}, with bijection $is_{MV}:\I_{MV} \simeq \W$ such that for any $v^* \in \I_{MV}$ we have that $I^*_{B} = is_{MV}(v^*):\L\rightarrow \mathfrak{Rm}$.
\end{definition}
 Now we can demonstrate that the many-valued semantics
given by Definition  \ref{def:MV-algebra} of the logic $\L_{in}$ is
corresponds to the two-step concept-algebra semantics given by
Definition \ref{def:MVintensemant}: \index{many-valuedness and concept-algebra}
\begin{theo} \label{theo:MV-algebra}\textsc{Many-valuedness and concept-algebra
semantics:}\\
For any fixed intensional interpretation $I$ of $\L_{in}$ into algebra of concepts $\mathcal{A}\mathfrak{m}_{int}$ there is a bijection $is_{in}:\W\simeq \E_{in}$, so that for any 4-valued Belnap-s interpretation $v^*\in \I_{MV}$, that is, MV-interpretation $I^*_{B} = is_{MV}(v^*)\in \W$ (in Definition \ref{def:MV-interpret}) the correspondent equivalent extensionalization  functions is $h = is_{in}(I^*_{B})\in \E_{in}$ (by T-constraint (\ref{eq:MVtarski}) and Definition \ref{def:built-in}), and viceversa, given any extensionalization homomorphism $h:\mathcal{A}\mathfrak{B}_{int} \rightarrow\A_{\mathfrak{Rm}}$ the correspondent equivalent MV-interpretation of $\L_{in}$ is $I^*_{B} = is_{in}^{-1}(h) \in
\W$, so that the following truth-diagram
\vspace*{-2mm}
\begin{diagram}
    &&   \mathcal{A}\mathfrak{B}_{int}  & \\
  &\ruTo^{I} &  & \rdTo^{h}  \\
\L_0 & & \rTo^{I^*_{B}} &  & \A_{\mathfrak{Rm}}
\end{diagram}
commutes, as a consequence of the basic Herbrand base correspondence
\begin{equation} \label{eq:groundEQ}
I^*_{B}(p_i^k(u_1,...,u_k)) = h(I(p_i^k(u_1,...,u_k)))
\end{equation}
for each ground atom $p_i^k(u_1,...,u_k)$ in Herbrand base $H \subset \L_0$.\\
 The  arrow $I^*_{B}:\L_0 \rightarrow\widetilde{X} = \mathfrak{Rm}_1 \subset \mathfrak{Rm}$ is the set-based Belnap's 4-valued truth semantics, while the  arrow  $h \circ I:\L_0\rightarrow \A_{\mathfrak{Rm}}$ corresponds to the two-step concept-algebra semantics with $I^*_B = h\circ I$.
\end{theo}
\textbf{Proof:} Notice that only the arrow $h$ in the diagram above
is a homomorphism (\ref{eq:int-ExtHomomorph}), \index{homomorphism of algebras}
 while $I$ is a non-homomorphic restriction of the homomorphism $I:\mathcal{A}\mathfrak{B}_{FOL} \rightarrow \mathcal{A}\mathfrak{B}_{int}$ given by (\ref{eq:IntensHomom}) in Corollary \ref{coro:intensemantMV} to the subset $\L_0$ of closed formulae of the free syntax algebra $\mathcal{A}\mathfrak{B}_{FOL}$ of the $\L_{in}$. Thus, this diagram is not homomorphic.
In what follows we consider that the intensional interpretation is fixed one.
We have to show that for any given  extensionalization  function $h$ the MV-interpretation $I^*_{B} = h \circ I$ is uniquely determined by the commutative diagram above, that is, from the fact that for any $\phi \in \L$ and assignment $g\in
\D^{\V}$ it holds that $v^*(\phi/g) = is_{MV}^{-1}(I_{B}^*)
(\phi/g)$, that is, $is_{in}^{-1}(is_{MV}(v^*))(\phi/g) = h (I(\phi/g))$.

First, from the correspondence (\ref{eq:groundEQ}), this diagram commutes for each ground atom in Herbrand base, and we need only to show that it is valid for any other composed sentence in $\L_0$. Notice that it is satisfied for the unary self-reference predicate $p_1^1$ and for the binary identity predicate $p_1^2$ as well. We will use the structural induction on number of logic operators of the sentences in $\L_0$. Suppose that it holds for every closed formula $\psi_1/g\in\L_0$ (with $R_1=Com(h(I(\psi/g))) = \{a_1\}$ and $a_1 = v^*(\psi_1/g)$) with $n-1$ connectives, and $\psi_2/g \in \L_0$ (with $R_2=Com(h(I(\psi/g))) = \{a_2\}$ and $a_2 = v^*(\psi_2/g$) with less than $n$ logic connectives, and let us demonstrate
that it holds for any formula $\phi$ with $n$ or more connectives.
There are the following cases:

1. $ \phi = \neg \psi_1$. Then, $h(I(\neg\psi_1/g)) = h(\widetilde{\neg} I(\psi_1/g)) =$ (by Def.\ref{def:Int-algebra}) $= \oslash h(I(\psi_1/g)) = $ (by Def. \ref{def:m-extensions})

$=  \left\{
    \begin{array}{ll}
 \{\neg a_1\}, & \hbox{if $~~ \neg a_1 \neq \bot$}\\
      \emptyset, & \hbox{otherwise}
    \end{array}
  \right.$, \\ while
 $I^*_{B}(\neg \psi_1/g) =$ (from (\ref{eq:mvR})) $= \left\{
    \begin{array}{ll}
 \{\neg v^*(\psi_1/g)\}=\{\neg a_1\}, & \hbox{if $~~ \neg a_1 \neq \bot$}\\
      \emptyset, & \hbox{otherwise}
    \end{array}
  \right.$.

2.  $ \phi =  (\psi_1 \wedge_S \psi_2)$. Then, $h(I(\phi/g)) = h(
I(\psi_1/g)\sqcap_S I(\psi_2/g)) =$ (by Def.\ref{def:Int-algebra}) $= h(
I(\psi_1/g))\otimes_S h(I(\psi_2/g))) = R_1\otimes_S R_2 =$ (by Definition \ref{def:m-extensions})

$= \left\{
    \begin{array}{ll}
 \{a_1\wedge a_2\}, & \hbox{if $~~ a_1\wedge a_2 \neq \bot$}\\
      \emptyset, & \hbox{otherwise}
    \end{array}
  \right.$,  \\while  $I^*_{B}(\psi_1/g \wedge_S
\psi_2/g) = $ (from Definition \ref{def:MV-algebra} and (\ref{eq:mvR}))

 $ = \left\{
    \begin{array}{ll}
\{v^*(\psi_1/g) \wedge v^*(\psi_2/g)\}= \{a_1\wedge a_2\}, & \hbox{if $~~ a_1\wedge a_2 \neq \bot$}\\
      \emptyset, & \hbox{otherwise}
    \end{array}
  \right.$\\
Analogosly we obtain for cases when $ \phi$ is  $\psi_1 \vee_S \psi_2$ or $
\psi_1 \Rightarrow_S \psi_2$.\\

3. $ \phi =  (\psi_1 \Leftrightarrow \psi_2)$. Let by inductive hypothesis $I^*_{B} (\psi_i/g) =h(I(\psi_i/g))$ for $i = 1,2$. Then,

 $h(I(\phi/g)) = h(
I(\psi_1/g)\leftrightarrow I(\psi_2/g)) =$ by Definition \ref{def:Int-algebra}

 $= \circledast (h(
I(\psi_1/g)), h(I(\psi_2/g)))$ by Definition \ref{def:m-extensions} and Corollary \ref{coro:intens}

$= \circledast (I^*_{B}
(\psi_1/g), I^*_{B}(\psi_2/g))$,  by inductive hypothesis

$= \circledast (\{v^*(\psi_1/g)\}, \{v^*(\psi_2/g)\})$  from (\ref{eq:mvR})

$= \otimes(Inc(v^*(\psi_1/g)\}, \{v^*(\psi_2/g)\}),Inc(v^*(\psi_2/g)\}, \{v^*(\psi_1/g)\}))$ by Def. \ref{def:m-extensions}

$= \{v^*(\psi_1/g)\} \Leftrightarrow v^*(\psi_2/g)\} \in \{\{f\},\{t\}\}$

$= \{v^*(\psi_1/g \Leftrightarrow \psi_2/g) \}$   from Definition \ref{def:MV-algebra}

  $= I^*_{B}(\psi_1/g \Leftrightarrow \psi_2/g) $  from (\ref{eq:mvR})

 $=I^*_{B}(\phi/g )$. \\

4.  $~~ \phi =  (\exists x_1)\psi(x_1)$, with $h(I(\psi(x_1))) = I_{B}(\psi(x_1))$, in accordance with (\ref{eq:MVtarski}) and (\ref{eq:openEQ}), and $R= h(I(\psi(x_1)))$.\\
Then, $h (I(\phi/g))= h (I((\exists x_1)\psi(x_1))) =
h( \exists_1 I(\psi(x_1))) =$ (by Def.\ref{def:Int-algebra})

  $= \boxplus_1 (h (I(\psi(x_1))))=$ (by Def. \ref{def:m-extensions})
$=  \left\{
    \begin{array}{ll}
 b = \bigvee\{a ~|~(u,a) \in R\}, & \hbox{if $~~ b \neq \bot$}\\
      \emptyset, & \hbox{otherwise}
    \end{array}
  \right.$, \\
 while $I^*_{B}(\phi/g) = I^*_{B}((\exists x_1)\psi(x_1))=$
 (from Def.\ref{def:MV-algebra} and (\ref{eq:mvR}))

 $=  \left\{
    \begin{array}{ll}
 b = \bigvee\{ a ~|~(u,a) \in R\}, & \hbox{if $~~ b \neq f$}\\
      \emptyset, & \hbox{otherwise}
    \end{array}
  \right.$\\
where  $\bigvee$ is the join operator in lattice $X$.
Analogously we obtain for cases when $\phi = (\forall x_i)\psi(x_1,...,x_i,...,x_k)$.
Thus, for any fixed intensional interpretation $I$ of $\L_{in}$, the diagram commutes and the bijection $is_{in}$ is well defined.
\\$\square$\\
This truth-diagram explains the equivalence of Belnap's 4-valued interpretations of the logic $\L_{in}$, and the two-step semantics for concept algebras. But it is given only for the set of ground (without free variables) formulae of logic $\L_0$, and its nucleus composed by
ground atoms of the Herbrand base $H$: the truth of (closed) sentences with quantifiers is uniquely determined by the truth of atoms of this Herbrand base.
\\\textbf{Remark}: Note that the logic of sentences $\L_0$ is based on the Belnap's truth-value algebra $\A_{B} = (X, \leq, \neg, \wedge, \vee,  \Rightarrow,\Leftrightarrow)$ in Definition \ref{def:mv-algebra} which is a modal Heyting algebra with unary modal paraconsistent negation operator $\neg$, so that the the sublogic $\L_0$ of $IFOL_B$ (composed only by sentence) is a kind of \emph{paraconsistent intuitionistic logic}, which is a constructive logic.
\\$\square$\\
 Now we will  extend this correspondence to open formulae in $\L$ as well.
\begin{coro} \textsc{Intensional/extensional Belnap's 4-valued  FOL semantics:} \label{coro:MVintalgebra}
\\For a given fixed intensional interpretation $I$ and any many-valued interpretation $v^*:\L_0 \rightarrow X$ of the $\L_{in}$, with MV-interpretation $I^*_{B} = is_{MV}(v^*)$, the following diagram, as a homomorphic extension of the diagram in Theorem \ref{theo:MV-algebra}, commutes,
\begin{diagram}
    &    &&& \mathcal{A}\mathfrak{B}_{int}~ (concepts/meaning) &&& &\\
 && \ruTo^{I ~(intensional~interpr.)}& & \frac{Frege/Russell}{semantics}  &&\rdTo^{h ~(extensionalization)} &&\\
 \mathcal{A}\mathfrak{B}_{FOL}~(syntax) & & &   & \rTo_{I^*_{B}~(MV-interpretation)} &&&& \A_{\mathfrak{Rm}} ~(denotation)   \\
\end{diagram}
where $h = is_{in}(I^*_{B})$ represents the bijective semantic
equivalence given by Theorem \ref{theo:MV-algebra} and the set of sentences $\L_0 \subset \L$ is substituted by the free syntax  algebra $\mathcal{A}\mathfrak{B}_{FOL}$ of the 4-valued many-sorted intensional first-order logic $\L_{in}$, introduced in Corollary \ref{coro:intensemantMV},  with the carrier set of all formulae $\L$ in the syntax algebra $\mathcal{A}\mathfrak{B}_{FOL}$.
\end{coro}
\textbf{Proof:} The homomorphism of intensional mapping $I$ is
defined by Corollary \ref{coro:intensemantMV}. Let us show that also for a given many-valued valuation $v^*\in  \I_{MV}$  which determines the MV-interpretation $I_{B}^* = is_{MV}(I_{mv})$ from Definition \ref{def:MV-interpret} and, from Theorem \ref{theo:MV-algebra}, the extensionalizzation $h = is_{in}(I^*_{B})$, the diagram above commutes.

 In fact, for any closed formula $\phi/g \in \L_0 \subset \L$, by definition of  $I_{B}^*$ we have that  $I_{B}^*(\phi) = $ (from the
commutative diagram in Theorem \ref{theo:MV-algebra}) $ = (h \circ
I)(\phi) = h(I(\phi))$, and the set of different "possible worlds"
\begin{equation}\label{eq:SetExtens}
\W = \{I^*_{B} = is_{MV}(v^*)~|~ v^* \in \I_{MV} \subset X^H\}~~\simeq \I_{MV}
\end{equation}
For any open formula (virtual predicate) $\phi(x_1,...,x_k) \in \L$,
from definition of  $I_{B}^*$ we have that
$I_{B}^*(\phi(x_1,...,x_k)) = \{(g(x_1),...,g(x_k),a)~|~g \in
\D^{\V}$ and $a = v^*(\phi/g)  \neq \bot\} = $(from Theorem
\ref{theo:MV-algebra}) $ = \{(g(x_1),...,g(x_k),a)~|~g \in \D^{\V}$
and $a = h(I(\phi/g)) \neq \bot\}
 = $ (from many-valued generalization of Tarski's constraint \cite{Majk22}
for the algebraic two-step semantics in Definition
\ref{def:MVintensemant}) $ =  h(I(\phi(x_1,...,x_k)))$.\\
Thus the diagram above commutes for every formula in $\L$, and we
obtain that $I_{B}^* = I \circ h$ is a \emph{many-valued
homomorphism} between the free syntax algebra of the logic $\L_{in}$
and the extensional algebra $\A_{\mathfrak{Rm}}$ of m-extensions.
\\$\square$\\
This homomorphic diagram formally express the fusion of Frege's and
Russell's semantics \cite{Freg92,Russe05,WhRus10} of meaning and
denotation of the FOL language, and renders mathematically correct
the definition of what we call an "intuitive notion of
intensionality", in terms of which a language is intensional if
denotation is distinguished from sense: that is, if both a
denotation and sense is ascribed to its expressions. This notion is
simply adopted from Frege's contribution (without its infinite
sense-hierarchy, avoided by Russell's approach where there is only
one meaning relation, one fundamental relation between words and
things, here represented by one fixed intensional interpretation
$I$), where the sense contains mode of presentation here described
algebraically as an algebra of concepts (intensions) $\mathcal{A}\mathfrak{B}_{int}$, and where sense determines denotation for any given extensionalization
function $h$ (corresponding to a given many-valued interpretation
$v^*$). More about the relationships between Frege's and
Russell's theories of meaning may be found in the Chapter 7,
"Extensionality and Meaning", in \cite{Beal82}.

As noted by Gottlob Frege and Rudolf Carnap (he uses terms
Intension/extension in the place of Frege's terms sense/denotation
\cite{Carn47}), the two logic formulae with the same denotation
(i.e., the same m-extension for a given many-valued interpretation
$v^*$) need not have the same sense (intension), thus such
co-denotational expressions are not \emph{substitutable} in general.

Often "intension" has been used exclusively in connection with
possible worlds semantics, however, here we use (as many others; as
Bealer for example) "intension" in a more wide sense, that is as an
\emph{algebraic expression} in the intensional algebra of meanings
(concepts) $\mathcal{A}\mathfrak{B}_{int}$ which represents the structural composition of more complex concepts (meanings) from the given set of atomic meanings.

So, not only the denotation (extension) is compositional, but also the meaning (intension) is compositional. So, the Montague's possible worlds representation \cite{Lewi86,Stal84,Mont70,Mont73,Mont74} the intension  can be generalized for the many-valued intensional FOL as well: \index{many-valued Montague's intension}
\begin{definition} \label{def:MVMontagueIntens} \textsc{MV-Montague's intension}:\\
 For a given many-valued intensional FOL language $\L_{in}$ with the set of logic formulae $\L$, the MV-Montague \emph{intension} as a mapping (higher-order function)
 $$I_n:\L\rightarrow \mathfrak{Rm}^{\W}$$
 where the set of possible worlds  $~\W$ is the set of MV-interpretations given by (\ref{eq:SetExtens}).
 Thus, for any virtual predicate $\phi(x_1,...,x_k)\in \L$ we obtain the mapping $f_k=I_n(\phi(x_1,...,x_k)):\W \rightarrow \mathfrak{Rm}$, such that for each $I^*_{B}\in \W$, $R = I_n(\phi(x_1,...,x_k))(I^*_{B}) \in \mathfrak{Rm}$  is equal to the relation $I^*_{B}(\phi(x_1,...,x_k))$ obtained from the MV-interpretation $I^*_{B}$ (derived from many-valued interpretation $v^* = is_{MV}^{-1}(I^*_{B}):\L_0\rightarrow X$ where the bijection $is_{MV}$ is defined in  Definition \ref{def:MV-interpret}).
 That is, for each virtual predicate $\phi(x_1,...,x_k)\in \L$ and "world" $h \in \W$, from the commutative diagram of Corollary \ref{coro:MVintalgebra}, the equation
 \begin{equation}\label{eq:MV-MontIntension}
  I_n(\phi(x_1,...,x_k))(I^*_{B}) = I^*_{B}(\phi(x_1,...,x_k))
 \end{equation}
  explains how works the idea of MV-Montague's intension.
\end{definition}
\section{ Kripke Semantics of $IFOL_B$\label{section:Kripke}}
 The distinction between intensions and extensions is important.
  So, we need also the actual world mapping
 which maps any intensional entity to its \emph{actual world
 extension}. As we have seen previously for the generalized  Montague's representation of the  intension in Definition \ref{def:MVMontagueIntens} for $\L_{in}$ (i.e., the 4-valued many-sorted intensional logic $IFOL_B$), we identified the \emph{possible worlds} $\W$ as a particular set of MV-interpretations and by a  particular mapping $I_n$ which assigns to intensional entities (corresponding to virtual predicates $\phi(x_1,...,x_k)\in \L$) their
 extensions in such possible world $w = I^*_{B} \in \W$, such that is satisfied the equation (\ref{eq:MV-MontIntension}). It is the direct bridge between
 intensional and the possible worlds representation
 \cite{Lewi86,Stal84,Mont70,Mont73,Mont74}, where intension of a proposition is a
 \emph{function} from a set of possible worlds $\W$ to truth-values, and
 properties and functions (in predicate intensional logics each functional $k$-ary symbol $f^k_i$ is substituted by a $(k+1)$-ary predicate) from $\W$ to sets of possible (usually  not-actual) objects.

  In propositional modal logics the possible worlds are entities where
a given propositional symbol can be true or false. Thus, from
\emph{logical} point of view the possible worlds in Kripke's
relational semantics are characterized by property to determine the
truth of logic sentences. The important question relative to the
syntax of the $\L_{in}$ is  if there is a kind of basic set of possible
worlds that have such properties. The answer is affirmative as in the case of the standard two-valued FOL, and we can use these results by adapting them to the many-valued approach.

In fact, if we consider a k-ary many-valued predicate letter $p_i^k$
as a new kind of "propositional letter", then an assignment
$g:\V_{X} \rightarrow \D$ can be considered as an \emph{intrinsic}
(par excellence) possible world, where the truth of this
"propositional letter" $p_i^k$ is equal to the truth of the ground
atom $p_i^k(g(x_1),...,g(x_{k}))\in \L_0$ of the many-valued intensional first-order logic $\L_{in}$. Here
\begin{equation}
\V_{X} = \V \bigcup \{x_{X}\}
\end{equation}
 where $x_{X}$ is the new typed "logic" variable  of the sort "\textbf{truth values}: s" introduced by Definition \ref{def:built-in} (whose domain is the set of Belnap's truth-values in $X$) not used in the set of variables $\V$ of $\L_{in}$  fixed in
Definition \ref{def:syntax}, so that for any assignment $g \in
\D^{\V_{X}}, g(x_{X}) \in X$.

 Consequently, in what follows,
analogously to the two-valued framework of the generalized Kripke semantics for predicate modal logics, given by Definition 12 in Section 1.2.1 in \cite{Majk22},
we will denote by $\W$ the set of \emph{explicit} possible worlds (defined
explicitly for each particular case of  modal logics, and in Definition \ref{def:MV-interpret} for this logic $\L_{in}$), while the set of assignments $\D^{\V_{X}}$  will be called "the set of \emph{intrinsic} possible worlds" (which is invariant and common for every \emph{predicate} modal logic).
By $\mathbb{W} \subseteq \{(w,g)~|~ w \in \W, g \in \D^{\V_{X}}\}$
we will denote the set of such (generalized) possible worlds. In this
way, as in the case of propositional modal logic, we will have that
a formula $\varphi$ is \emph{true} in a Kripke's interpretation
${\M}$ if  for each (generalized) possible world $u = (w,g) \in
\mathbb{W}$, $~~{\M} \models_{u}~\varphi$.

We denote by $\|\phi\| = \{(w,g) \in \mathbb{W}~|~\M \models_{w,g}~\phi\}$ the set of all worlds where the  formula $\phi$ is satisfied by interpretation $\M$. Thus, as in the case of propositional modal logics, also in the case of predicate modal logics we have that a formula $\phi$ is true iff  it is satisfied in all (generalized) possible worlds, i.e., iff $\|\phi\| = \mathbb{W}$.

With this new arrangement we can reformulate the generalized Kripke semantics
for two-valued modal predicate logic (as usual, $\pi_1$ and $\pi_2$ denote the first and the second projections, and we recall that we replaced in intensional FOL each functional symbol by corresponding predicate symbol, so the set $F$ of functional letters is empty):
\begin{definition} \label{def:MVKripSemG} \textsc{Generalized Kripke semantics
for $\L_{in}$}:\\
 We denote by $\M = (\mathbb{W}, \{$$
{\R}_i \}, \D, I_K)$ a multi-modal  Kripke's interpretation with
 a set of (generalized) possible worlds $\mathbb{W} = \W\times \D^{\V_{X}}$, a set of explicit possible worlds $\W = \pi_1(\mathbb{W})$
 and $ \pi_2(\mathbb{W}) = \D^{\V_{X}}$, the accessibility relations ${\R}_i \subseteq \mathbb{W} \times  \P(\mathbb{W})$,  $i = 1,2,...$,
 non empty domain $\D$, and   a function $~~I_K:\W\times P \rightarrow ~{\bigcup}_{n \in \N} \textbf{2}^{\D^n}$, such that for any explicit world $w \in \W$:
 For any predicate letter $p_i^k \in P$, the function $~I_K(w,p_i^k):\D^{k}\rightarrow \textbf{2}~$ defines the m-extension of $p_i^k$ in an explicit world $w$,

$~\|p_i^k(x_1,...,x_k)\|_{\M, w} =_{def} \{  (d_1,...,d_{k+1}) \in \D^{k}\times X~|~~I_K(w,p_i^k)(d_1,...,d_{k+1}) = t$ and $d_{k+1} \neq \bot \}$.
 \end{definition}
 Based on this generalized Kripke semantics for $\L_{in}$,  we can  present the Kripke semantics for 4-valued many-sorted  intensional first-order logic $\L_{in}$:\index{two-step intensional semantics}
 \begin{definition} \label{def:MVKripSem} \textsc{Kripke semantics for
 $\L_{in}$:}\\
 Let $(I,\W)$ be the  two-step intensional semantics in (\ref{eq:2-step}) for logic $\L_{in}$  (where the functional letters with  arity greater than zero are substituted by predicate letters) given in Definition \ref{def:MVintensemant} with $I^*_{B}\in \W$ possible worlds in (\ref{eq:SetExtens}).\\
 By $~\M = (\mathbb{W}, {\R}_{\neg}, {\R}_{\wedge}, {\R}_{\vee}, {\R}_{\Rightarrow}, {\R}_{\Leftrightarrow}, \{{\R}_{\exists x_i}, {\R}_{\forall
x_i}\}_{i \in \mathbb{N}}, \D, I_K)$ we denote a multi-modal Kripke's
interpretation with  a set of (generalized) possible worlds $\mathbb{W}$, a set of explicit possible worlds $\W = \pi_1(\mathbb{W})$  and $ \pi_2(\mathbb{W}) = \D^{\V_{X}}$, non empty domain $\D$, with the variable $x_{X}\in \D^{\V_{X}}$ is the typed "logic" variable  of the built-in sort "\textbf{truth values}: s" (concept in $D_2$ in Definition \ref{def:built-in} such that for $g\in \D^{\V_{X}}$, and any extensionalization function $h$, $g(x_X) \in \pi_1(h(\textbf{truth values}: s)) =X$ ), the accessibility  relations:
 \begin{description}
   \item[a.] ${\R}_{\neg} = \{(g,g_1)~|~g,g_1\in \D^{\V_{X}}$, such that for all $x \in \V, g_1(x) = g(x)$ and  $g(x_{X}) = \neg g_1(x_{X})\}$,
   \item[b.] ${\R}_{\odot} = \{(g,g_1,g_2)~|~g,g_1,g_2\in \D^{\V_{X}}$, such that for all $x \in \V, g_1(x) = g_2(x) = g(x)$ and
 $g(x_{X}) =  g_1(x_{X})\odot g_2(x_{X})\}$, for each 4-valued operation $\odot \in \{\wedge, \vee, \Rightarrow, \Leftrightarrow \}$ in the Belnap's truth-lattice $X$,
   \item[c.] ${\R}_{\odot} \subseteq \D^{\V_{X}} \times \P(\D^{\V_{X}})$,
 such that\\
 $(g,G) \in {\R}_{\odot}~$ iff $~ g(x_{X}) = \diamond \{g'(x_{X})~|~g' \in
 G$ and for all $x \in \V\backslash \{x_i\}, g'(x) = g(x) \}$,\\
 for each pair $(\odot, \diamond) \in \{(\exists x_i, \bigvee), (\forall x_i, \bigwedge)\}$, where $\bigwedge$ and $\bigvee$ are the meet and join operator of Belnap's truth-lattice $X$, respectively.
 \end{description}
 with mapping $I_K:\W\times P \rightarrow {\bigcup}_{n \in \N}
\textbf{2}^{\D^n\times X}$, such that for any explicit world $w=I^*_{B} \in \W$, i.e., many-valued interpretation $v^* = is_{MV}^{-1}(I^*_{B})$,
$p_i^k \in P$ and $(u_1,...,u_{k+1}) \in \D^{k}\times X$,

$~I_K(I^*_{B},p_i^k)(u_1,...,u_{k+1}) = t ~~$ iff $~~u_{k+1} =
v^*(p_i^k(u_1,...,u_{k}))$.\\
Then, for any world $(w,g) \in \mathbb{W}$ with $w = I^*_{B} \in \W$ and $g\in \D^{\V_{X}}$, we define the
many-valued satisfaction,  denoted by $~ \models_{w,g} ~$, as follows:

1. $~~\M\models_{w,g} p_i^k(x_1,...,x_k) ~~$ iff $~~I_K(w,p_i^k)(g(x_1),...,g(x_k),g(x_{X})) = t$,

2. $~~\M \models_{w,g} ~\neg \phi  ~~$ iff   $~~\exists g_1
 \in \D^{\V_{X}}.((g,g_1) \in {\R}_{\neg}$ and $~~\M \models_{w,g_1}~\phi)$,

3. $~~\M \models_{w,g} ~\phi \circledast \psi ~~$\\ iff $~~\exists g_1,g_2
 \in \D^{\V_{X}}.((g,g_1,g_2) \in {\R}_{\odot}$ and $~~\M
\models_{w,g_1}~\phi$ and $~~\M \models_{w,g_2} ~\psi$), \\$~~$ for each FOL algebra syntax operator $\circledast \in \{\wedge_S, \vee_S, \Rightarrow_S, \Leftrightarrow \}$  with $S$ set of pairs of indices of equal variables in virtual predicates $\phi$ and $\psi$, and 4-valued operator in Belnap's truth lattice $\odot \in \{\wedge, \vee, \Rightarrow, \Leftrightarrow\}$, relatively,

4. $~~\M \models_{w,g}~ ~(\exists x_i) \phi(x_1,...,x_i,...,x_k)~~$\\
iff $~~$ exists $(g,G) \in {\R}_{\exists x_i}$  such that $~G = \{g_1\in \D^{\V_{X}}~|~\forall_{1\leq j\leq k} j\neq i. (g_1(x_j) = g(x_j))$ and $g_1(x_X) \in Com(w(\phi/g_1))\}$,     and  $(g_1 \in G~$ iff $~\M\models_{w,g_1} ~\phi$),

5. $~~\M \models_{w,g}~ ~(\forall x_i) \phi(x_1,...,x_i,...,x_k)~~$\\
iff $~~$ exists $(g,G) \in {\R}_{\forall x_i}$  such that $~G = \{g_1\in \D^{\V_{X}}~|~\forall_{1\leq j\leq k} j\neq i. (g_1(x_j) = g(x_j))$ and $g_1(x_X) \in Com(w(\phi/g_1))\}$,     and  $(g_1 \in G~$ iff $~\M\models_{w,g_1} ~\phi$),
 \end{definition}
 Let us show that this multi-modal Kripke semantics of $\L_{in}$ is correct
 semantics, able to support the Montague's approach to the  intensional semantics.
\begin{theo} For any given multi-modal Kripke interpretation $\M$ of
$\L_{in}$, it is valid that for any formula $\phi(x_1,...,x_k) \in\L, k \geq 0$, and world $(w,g) \in \mathbb{W}$ where $w = I^*_{B}$ with many-valued interpretation $v^* = is_{MV}^{-1}(w)$:
\begin{equation} \label{eq:MVmodel}
~~\M \models_{w,g}~ ~ \phi(x_1,...,x_k)~~~~ iff ~~~~g(x_{X}) =
v^*(\phi(x_1,...,x_k)/g)
\end{equation}
that is, by Theorem \ref{theo:MV-algebra}, iff $~~( g(x_{X})\neq \bot$
and $w(\phi(x_1,...,x_k)/g) = \{g(x_{X})\}$,\\ or $g(x_{X}) =
\bot$ and $w(\phi(x_1,...,x_k)/g) = \emptyset$). \\
Consequently, the m-extension of $\phi(x_1,...,x_k)$ computed by the
multi-modal Kripke's interpretaion $\M$ in a given explicit world $w= I^*_{B}
\in \W$ is equal to m-extension of the same formula computed by the
 intensional semantics of the $\L_{in}$, that is,
 \begin{equation} \label{eq:MVmodel1}
~~\|\phi(x_1,...,x_k)\|_{\M, w} = w(\phi(x_1,...,x_k))
\end{equation}
Thus, from the Montague's approach to the intension (meaning) of the logic
formulae given by the mapping $I_n: \L \rightarrow
\mathfrak{Rm}^{\W}$ in Definition \ref{def:MVMontagueIntens}, it holds that
$I_n(\phi(x_1,...,x_k))(w) = \|\phi(x_1,...,x_k)\|_{\M, w}$ is the m-extension of this formula in this possible world. Consequently, this multi-modal Kripke semantics for $\L_{in}$ is adequate to support Montague's intensional semantics.
\end{theo}
 \textbf{ Proof:} Let us prove (\ref{eq:MVmodel}) by structural induction on number of logic  operators in a virtual predicate (open formula) $\phi(x_1,...,x_k)$:

 1. Let $\phi(x_1,...,x_k)$ be an atom $p_i^k(x_1,...,x_k)$ (with
 zero logic operators). Consequently, if $~\M \models_{w,g}~ ~ p_i^k(x_1,...,x_k)$
 then, from point 1 of Definition \ref{def:MVKripSem} we have that

 $~I_K(w,p_i^k)(g(x_1),...,g(x_k),g(x_{X})) = t$, that is $g(x_{X}) = v^*(p_i^k(x_1,...,x_k)/g)$. And vice versa.

 2. Let us suppose that this theorem holds for each open formula
 with less than $n\geq 1$ logic operators, and consider any formula
 $\phi(x_1,...,x_k)$ with $n$ operators. There are the following  cases:
 2.1. $\phi(x_1,...,x_k) = \neg \psi (x_1,...,x_k)$ where  $\psi
 (x_1,...,x_k)$ has less than $n$ operators. Let  $~\M \models_{w,g}
 \phi(x_1,...,x_k)$, so that from the point 2 of Definition
 \ref{def:MVKripSem} we obtain that $~\M \models_{w,g_1}
 \psi(x_1,...,x_k)$ with $ g_1(x_{X}) = a, g(x_{X}) = \neg a$, and
 consequently, by inductive hypothesis  $v^*(\psi(x_1,...,x_k)/g_1)
 = v^*(\psi(x_1,...,x_k)/g) =  a$.
Then, $g(x_{X}) = \neg a = \neg v^*(\psi(x_1,...,x_k)/g) = $
(from point 1 of Definition \ref{def:MV-algebra}) $ = v^*(\neg
\psi(x_1,...,x_k)/g) = v^*(\phi(x_1,...,x_k)/g))$. And viceversa. Analogously for another unary logic connectives different from the quantifiers.

 2.2. $\phi(x_1,...,x_k) =  \psi_1 \wedge_S \psi_2$ where  $\psi_1, \psi_2$ have less than $n$ operators.\\ Let  $~\M \models_{w,g}
 \phi(x_1,...,x_k)$, so that from the point 3 of Definition
 \ref{def:MVKripSem} we obtain that $~\M \models_{w,g_1}
 \psi_1$  and $~\M \models_{w,g_2}
 \psi_2$ with $ g(x_{X}) = g_1(x_{X}) \wedge g_2(x_{X})$.
 Consequently, by inductive hypothesis  $g_1(x_{X}) = v^*(\psi_1/g_1)
 = v^*(\psi_1/g))$ and $g_2(x_{X}) = v^*(\psi_2/g_2)
 = v^*(\psi_2/g)$.
 Then, $g(x_{X}) =  g_1(x_{X}) \wedge g_2(x_{X}) = v^*(\psi_1/g) \wedge  v^*(\psi_1/g) =$
(from point 2 of Definition \ref{def:MV-algebra}) $ = v^*(
\psi_1/g \wedge \psi_2/g) = v^*(\phi(x_1,...,x_k)/g)$.
 And viceversa. Analogously we can show that it holds for the cases
 when $\phi(x_1,...,x_k)$ is $\psi_1 \odot \psi_2$ for other binary connectives as well.

 2.3. $\phi(x_1,...,x_k) =  (\exists x_i)\psi$ where  $\psi$ has less than $n$ operators.\\ Let  $~\M \models_{w,g} \phi(x_1,...,x_k)$, so that from the point 4 of Definition  \ref{def:MVKripSem} we obtain that $~\M \models_{w,g}
 (\exists x_i)\psi$  with $ g(x_{X}) = \bigvee G = \bigvee \{g_i(x_{X}) ~| ~\M \models_{w,g_i} \psi$, $g_i \in \D^{\V}$ such that for all $x \in \V \backslash \{x_i\}, g_i(x) =  g(x)\}$, so that from the  inductive hypothesis  $ g(x_{X}) = \bigvee \{g_i(x_{X})~| ~g_i(x_{X}) = v^*(\psi/g_i)$, $g_i \in \D^{\V}$ such that for all $x \in \V \backslash \{x_i\}, g_i(x) = g(x)\} = \bigvee \{v^*(\psi/g_i)~| ~g_i \in \D^{\V}$ such that for all $x \in \V \backslash \{x_i\}, g_i(x) =
 g(x)\}=$ (from point 3 of Definition \ref{def:MV-algebra}) $
 = v^*((\exists x_i)\psi)$. And viceversa.\\
  Analogously we can show that it holds for the case
 when $\phi(x_1,...,x_k) = (\forall x_i)\psi$ as well.\\
 Consequently we obtain (\ref{eq:MVmodel1}),\\ $\|\phi(x_1,...,x_k)\|_{\M, w} =_{def} \{(g(x_1),...,g(x_k),g(x_{X}))~|~g \in \D^{\V}$ and $\M \models_{w,g}
 \phi(x_1,...,x_k) \} = w(\phi(x_1,...,x_k))= I^*_{B}(\phi(x_1,...,x_k))=  I_n(\phi(x_1,...,x_k))(I^*_{B})$,  from (\ref{eq:MV-MontIntension}),\\ and hence correspondence with Montague's approach to the intension (meaning).
 \\$\square$ \\
 Notice that the meaning of the logic formulae is given by mapping
$I_n: \L \rightarrow \mathfrak{Rm}^{\W}$, such that for any formula $\phi(x_1,...,x_k) \in \L$ its Montague's meaning is a function $I_n(\phi(x_1,...,x_k)):\W \rightarrow R$  that maps each explicit possible world into the m-extension of this formula in each given possible world, given by

$I_n(\phi(x_1,...,x_k))(w) = \|\phi(x_1,...,x_k)\|_{\M, w}$.\\
This meaning is \emph{compositional}. That is, the meaning of a complex logic
formula (or the complex concept obtained by applying the intensional
interpretation $I$ to this formula), is functionally dependent on
meanings of its subformulae. It is result of the fact that for any formula,

 $I_n(\phi(x_1,...,x_k))(w) = \|\phi(x_1,...,x_k)\|_{\M, w} = w(\phi(x_1,...,x_k))
= I^*_{B}(\phi(x_1,...,x_k))$,\\
from the fact that $w = I^*_{B}$ and is a \emph{homomorphism} between free syntax algebra $\mathcal{A}\mathfrak{m}_{FOL}$ and algebra of m-extensions $\A_{\mathfrak{Rm}}$ given in Corollary \ref{coro:MVintalgebra}. For example,

$I_n(\phi\wedge_S \psi)(w) = I^*_{B}(\phi\wedge_S \psi) = I^*_{B}(\phi) \otimes_S  I^*_{B}(\phi) = I_n(\phi)(w) \otimes_S I_n(\psi)(w)$, \\
where $S$ is uniquely determined by the common subset of free variables in $\phi$ and in $\psi$.\\
Consequently, the Kripke semantics of $\L_{in}$ given by Definition
\ref{def:MVKripSem}, based on possible worlds and accessibility
relations, represents this intensional aspect of the logic $\L_{in}$.\\
The meaning, concept-algebra semantics and the Belnap's 4-valuedness of any
logic formulae  $\phi(x_1,...,x_k)$ in $\L_{in}$, can be represented
synthetically by the following equalities:\\\\
$I_n(\phi(x_1,...,x_k))(w)$ (Montague's "meaning" semantics)

$ = \|\phi(x_1,...,x_k)\|_{\M, w} $ (Kripke semantics)

$ = I^*_{B}(\phi(x_1,...,x_k)) $ (concept-algebra semantics)

$ = \{(g(x_1),...,g(x_k),v^*(\phi(x_1,...,x_k)/g))~|~g \in
\D^{\V} \}$ (many-valued semantics),\\\\
where $I^*_{B} = is_{in}^{-1}(h)$ (from Corollary \ref{coro:MVintalgebra}) is a MV-interpretation of $\L_{in}$, equivalent to the Bealer's concept-extensionalization function $h$ for a given fixed intensional homomorphic interpretation $I:\mathcal{A}\mathfrak{m}_{FOL} \rightarrow \mathcal{A}\mathfrak{m}_{int}$.

\section{Conclusions}
%
This paper is a continuation of  my approach to AGI (Strong-AI) for a new generation of intelligent robots using \emph{natural languages} based on neuro-symbolic AI attempts to integrate neural and symbolic architectures in a manner that addresses strengths and weaknesses of each, in a complementary fashion, in order to support robust strong AI capable of reasoning, learning, and cognitive modeling. In order to provide a symbolic architecture of modern robots, able to use natural languages to communicate with humans and to reason about their own knowledge with self-reference and abstraction language property. If progress in theoretical neuroscience continues, it should become possible to tie psychological to neurological explanations by showing how mental representations such as \emph{concepts} are constituted by activities in neural populations, and how computational procedures such as spreading activation among concepts are carried out by neural processes. Concepts, which partly correspond to the words in spoken and written language, are an important kind of mental representation.

The Belnap's 4-valued bilattice of truth values is introduced in order to provide the AGI robots with human ability to consider also unknown and contradictory knowledge (as Liar paradoxes with logic truth value $\top$, for example ).\\
Future work will be dedicated to the knowledge properties of AGI robots, based on this Intensional FOL over Belnap's 4-valued bilattice of the truth-values.


%

\end{document}